\documentclass[11pt,mf,harvard]{article}
\usepackage{amssymb}
\usepackage{stmaryrd}
\usepackage{color}
\usepackage{epsfig}

\usepackage{hyperref}

\usepackage[
backend=biber,
style=apa,
sorting=nyt
]{biblatex}
\DeclareSourcemap{
  \maps[datatype=bibtex]{
    \map{
      \step[fieldset=issn, null]
      \step[fieldset=doi, null]
      \step[fieldset=url, null]
      \step[fieldset=urldate, null]
    }
  }
}

\hypersetup{
    colorlinks=true,
    urlcolor=blue,
    linkcolor=blue,
    citecolor=blue
}

\begin{document}
\begin{center}
{\Large\bf Inference in Incomplete Models}\\
\vskip30pt {\large Alfred Galichon and Marc Henry}

\vspace{12pt}

Harvard University and Columbia University \vskip5pt First draft:
September 15, 2005\\This draft\footnotemark[1]: May 26, 2006
\end{center}

\vskip30pt \footnotetext[1]{This research was carried out while
the first author was visiting the Bendheim Center for Finance,
Princeton University and financial support from NSF grant SES
0350770 to Princeton University and from the Conseil G\'en\'eral
des Mines is gratefully acknowledged. The authors also wish to
thank Gary Chamberlain, Xiaohong Chen, Victor Chernozhukov,
Pierre-Andr\'e Chiappori, Ronald Gallant, Peter Hansen, Han Hong,
Guido Imbens, Michael Jansson, Massimo Marinacci, Rosa Matzkin,
Francesca Molinari, Ulrich Mueller, Alexei Onatski, Ariel Pakes,
Jim Powell, Peter Robinson, Bernard Salani\'e, Thomas Sargent,
Jos\'e Scheinkman, Jay Sethuraman, Azeem Shaikh, Chris Sims,
Kyungchul Song and Edward Vytlacil and seminar participants at
Berkeley, Columbia, \'Ecole polytechnique, Harvard, MIT, NYU,
Princeton, SAMSI and Stanford for helpful comments (with the usual
disclaimer). Correspondence address: Department of Economics,
Columbia University, 420 W 118th Street, New York, NY 10027, USA.
mh530@columbia.edu. This paper is now superseded by various papers by the same authors.}

\begin{abstract}
We provide a test for the specification of a structural model
without identifying assumptions. We show the equivalence of
several natural formulations of correct specification, which we
take as our null hypothesis. From a natural empirical version of
the latter, we derive a Kolmogorov-Smirnov statistic for Choquet
capacity functionals, which we use to construct our test. We
derive the limiting distribution of our test statistic under the
null, and show that our test is consistent against certain classes
of alternatives. When the model is given in parametric form, the
test can be inverted to yield confidence regions for the
identified parameter set. The approach can be applied to the
estimation of models with sample selection, censored observables
and to games with multiple equilibria.
\end{abstract}

\vskip6pt \noindent {\scriptsize JEL Classification: C10, C12,
C13, C14, C52, C61
}

\newpage
\section*{Introduction}
In many contexts, the ability of econometric models to identify,
hence estimate from observed frequencies, the distribution of
residual uncertainty often rests on strong prior assumption that
are difficult to substantiate and even to analyze within the
economic decision problem.\vskip4pt

A recent approach, pioneered by Manski has been to forego such
prior assumptions, thus giving up the ability to identify a single
probability distribution for residual uncertainty, and allow
instead for a set of distributions compatible with the empirical
setup. A variety of models have been analyzed in this way, whether
partial identification stems from incompletely specified models
(typically models with multiple equilibria) or from structural
data insufficiencies (typically cases of data censoring). See
\cite{Manski:2005} for an up-to-date survey on the
topic.\vskip4pt

All these models with incomplete identification share the basic
fundamental structure that the residual uncertainty and the
relevant observable quantities are linked by a many-to-many
mapping instead of a one-to-one mapping as in the case of
identification.\vskip4pt

In this paper, we propose a general framework for conducting
inference without additional assumptions such as equilibrium
selection mechanisms necessary to identify the model (i.e. to
ensure that the many-to-many mapping is actually one-to-one). The
usual terminology for such models is ``incomplete'' or ``partially
identified.''\vskip4pt

In a parametric setting, the objective of inference in partially
identified models is the estimation of the set of parameters
(hereafter called {\em identified set}) which are compatible with
the distribution of the observed data and an assessment of the
quality of that estimation. For the latter objective, two routes
have been taken.\vskip4pt

\cite{CHT:2002} initiated research to obtain regions that
cover the identified set with a prescribed probability. They
propose an M-estimation approach with a sub-sampling procedure to
approximate quantiles of the supremum of the criterion function
over the identified set. \cite{Shaikh:2005} proposes an
alternative M-estimation with subsampling procedure that nests the
\cite{CHT:2002} proposal. M-estimation with subsampling is
the only general proposal to date that does not rely on a
conservative testing procedure, but the choice of criterion
function in the M-estimation procedure is arbitrary, and may have
a large effect on the confidence regions. \vskip6pt

In related research, a more direct application of random set
methods has been taken to achieve the goal of constructing
confidence regions for the identified set: \cite{SV:2005}
consider a special model where the identified set is a
deterministic mapping of a collection of expectations, and base
inference on the sample analogs of these expectations.
\cite{BM:2006} propose the use of central limit theorems for
random sets to conduct inference in models with set valued data.
However, the adaptation of delta theorems for random sets
is required for this approach to attain its full
potential.\vskip4pt

The second route was initiated by \cite{IM:2004} who
considered the different problem of covering each element of the
identified set, and demanded uniform coverage.
\cite{Shaikh:2005} shows that the M-estimation with
subsampling procedure can also be applied to uniform coverage of
elements of the identified set. \cite{PPHI:2004} consider
models that are defined by moment inequalities and propose a
conservative procedure to form a confidence region for all
parameters in the identified set based on inequalities testing
ideas.
The procedure is conservative since the limiting distribution of
the test statistic depends on the number of constraints that are
actually binding, and unlike in the special one dimensional
treatment response case analyzed by \cite{IM:2004}, no
superefficient pre-test is available.\vskip4pt

Still in the latter spirit, \cite{ABJ:2004} consider entry
games (and more generally games with discrete strategies) and
propose a conservative procedure to form a confidence region for
all parameters in the identified set based on the idea that the
probability of a certain outcome is no larger than the probability
that necessary conditions (such as Nash rationality constraints)
are met.\vskip4pt

The inference procedure proposed here is in the same spirit as
this latter contribution, but it gives a full formalization of the
idea in a very general framework, does not restrict the class of
distributions of observables (hence allows estimation of games
with continuous strategies as well as entry games), does not rely
on resampling procedures (though they may be used as alternative
quantile approximation devices), and provides an exact test as
opposed to the conservative procedures considered above.\vskip4pt

After a prelude to expound the ideas developed here in the
familiar case of Kolmogorov-Smirnov specification testing, the
general set-up is described (with some examples) in section~1. It
comprises the specification of a structure (in the Koopmans
terminology) with observable and unobservable variables
(unobservable to the analyst but not necessarily to the economic
agents) related by a many-to-many mapping as opposed to the
one-to-one mapping required for identification. The structure is
defined by the many-to-many mapping (which can comprise
rationality constraints as before, as well as any constraints that
are plausible within the theory) and a hypothesized distribution
for the unobserved variables. To fix ideas, we call $\Gamma$ the
many-to-many mapping defining the structure, $\nu$ a hypothesized
distribution of unobservables and $P$ the true distribution of
observables.\vskip4pt

Still in section~1, a characterization is given of what we mean by
correct specification, viz. compatibility of the structure with
the distribution of the observable variables, and it is shown that
several natural ways of defining compatibility are in fact
equivalent. They include (among other notions) a compatibility
notion based on selections $\gamma$ of $\Gamma$ (i.e. functions
such that $\gamma\in\Gamma$), a notion based on the existence of a
joint probability that admits $\nu$ and $P$ as marginals and is
supported on the region where the constraints implied by $\Gamma$
are satisfied, and the notion of maximum plausibility introduced
by \cite{Dempster:67}.\vskip4pt

Second, in section~2, we show that the characterizations of
correct specification of the structure are equivalent to the
existence of a zero cost solution to a Monge-Kantorovich mass
transportation problem, where mass is transported between
distribution $P$ and distribution $\nu$ with zero-one cost
associated with violation of the constraints implied by $\Gamma$.
This is the topic of section~2. Note that a special case of
Monge-Kantorovich transportation problem is the well-know matching
problem.\vskip4pt

Third, still in section~2, this observation allows us to conduct
inference using the empirical version of the mass transportation
problem (with the unknown $P$ replaced by the empirical
distribution $P_n$). Empirical formulations pertaining to the
different characterizations of correct specification of the
structure are compared, and several are found to be equivalent,
whereas others differ according to the choice of probability
metric. It turns out that the dual of the empirical problem yields
a statistic that reduces to the familiar Kolmogorov-Smirnov
specification test statistic in the identified case where $\Gamma$
is one-to-one.\vskip4pt

The properties of this statistic are examined in section~3. The
classical Kolmogorov-Smirnov statistic tests the equality of two
probability measures by checking their difference on a {\em good}
class of sets (large enough to be convergence-determining, but
small enough to allow asymptotic treatment). Here our test
statistic checks that $P(A)$ is no larger than $\nu(\Gamma(A))$
for all $A$ in a similar class of sets. Since $\nu(\Gamma(A))$ is
the probability of the sufficient conditions implied by $A$, we
see the strong similarity with the \cite{ABJ:2004} approach.
Hence the dual empirical problem provides us with a computable
test statistic, and a distribution to compare it to, and a
parallel with the classical case. \vskip4pt

We derive the asymptotic distribution of our test statistic and
describe how classes of alternatives against which our test has
power are related to what we call core-determining classes of
sets.

Finally, the fourth section shows simple implementation
procedures, and the inversion of the test to construct a
confidence region for the elements of the identified set of
parameters when both $\Gamma$ and $\nu$ are specified in
parametric form. If one is interested in testing structural
hypotheses such as extra constraints implied by theory, within the
framework of a partially identified model, the constraints should
be rejected if the region they imply on the parameter set does not
intersect with the identified set. Here the question can be
answered directly by incorporating the extra constraints in the
model and testing the restricted specification. If, on the other
hand, one is interested in reporting parameter value estimates
with confidence bounds for policy analysis, the specification test
can be inverted to the end of providing confidence regions that
cover the elements of the identified set with pre-determined
probability, or confidence regions that cover the identified set
itself.\vskip4pt

At the end of this section, we discuss semi-nonparametric
extensions of our approach to include models which do not specify
a parametric family of hypothesized data generating processes for
the unobservable variables. This includes as a special case models
defined by moment inequalities, the full treatment of which is the
subject of the companion paper \cite{GH:2006b}.\vskip4pt

The last section of the main text concludes; whereas proofs and
additional results are collected in the appendix.\vskip4pt

\section*{Prelude: complete model benchmark}
Before we define incomplete model specifications, we give a short
heuristic univariate description of the benchmark that we use and
discuss the Kolmogorov-Smirnov specification test statistic that
we are effectively generalizing in this paper.\vskip4pt

\begin{figure}[htbp]
\begin{center}
\includegraphics[width=8cm]{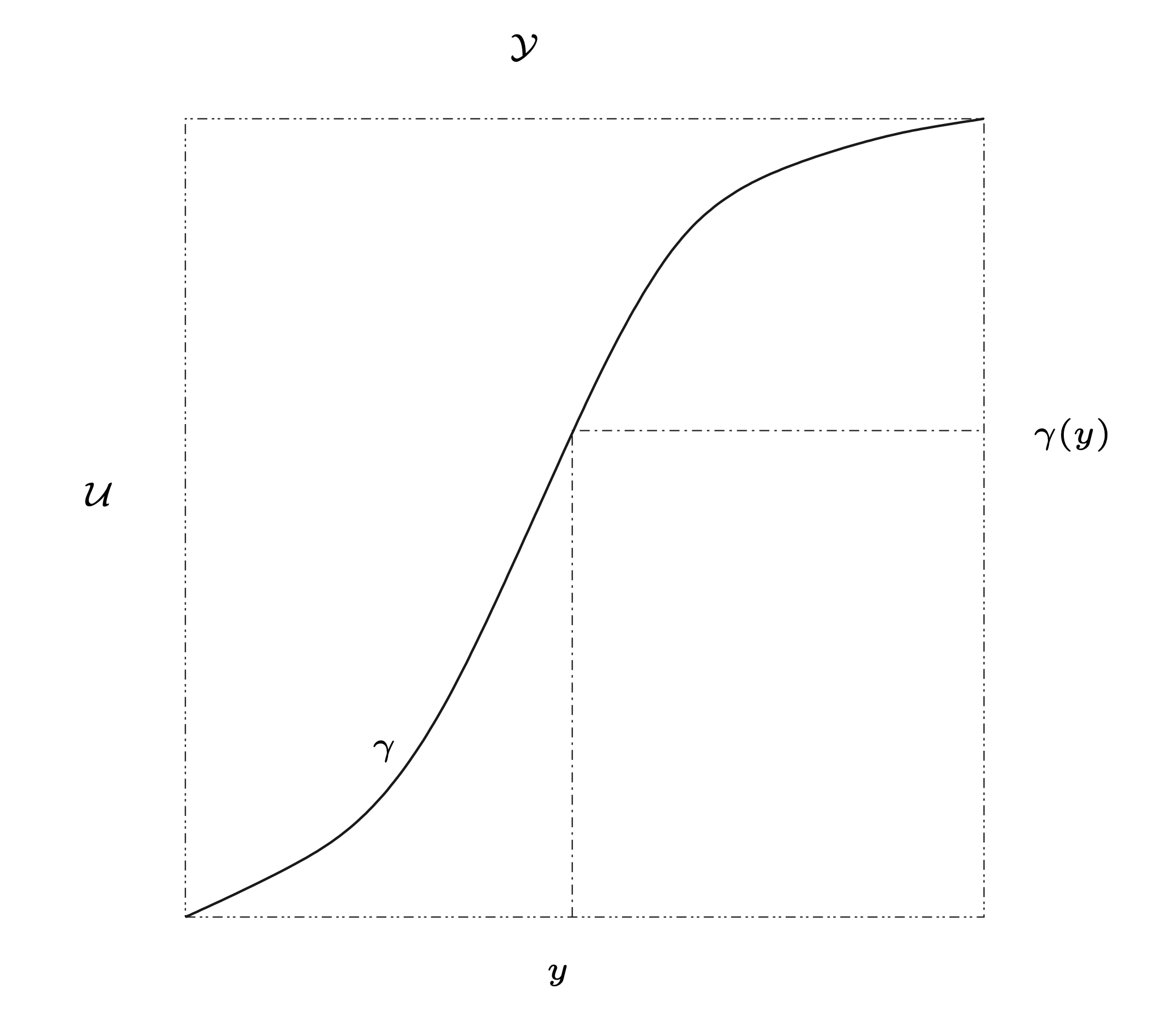}
\caption{Bijective structure}
\label{bijection}
\end{center}
\end{figure}

For ease of noptation, we consider observables $y\in\mathbb{R}$
and unobservables $u\in\mathbb{R}$ (also called ``unobserved
shocks'', ``latent variables'', etc...). Abstracting from
dependence on an unknown deterministic parameter, we define a
``complete'' structure as a pair $(\nu,\gamma)$, where $\nu$ is a
data generating process for the unobservables, and $\gamma$ is a
bijection from the set of observables to the set of unobservables,
as in figure~\ref{bijection}.\vskip4pt

\vskip6pt

\begin{figure}[htbp]
\begin{center}
\includegraphics[width=8cm]{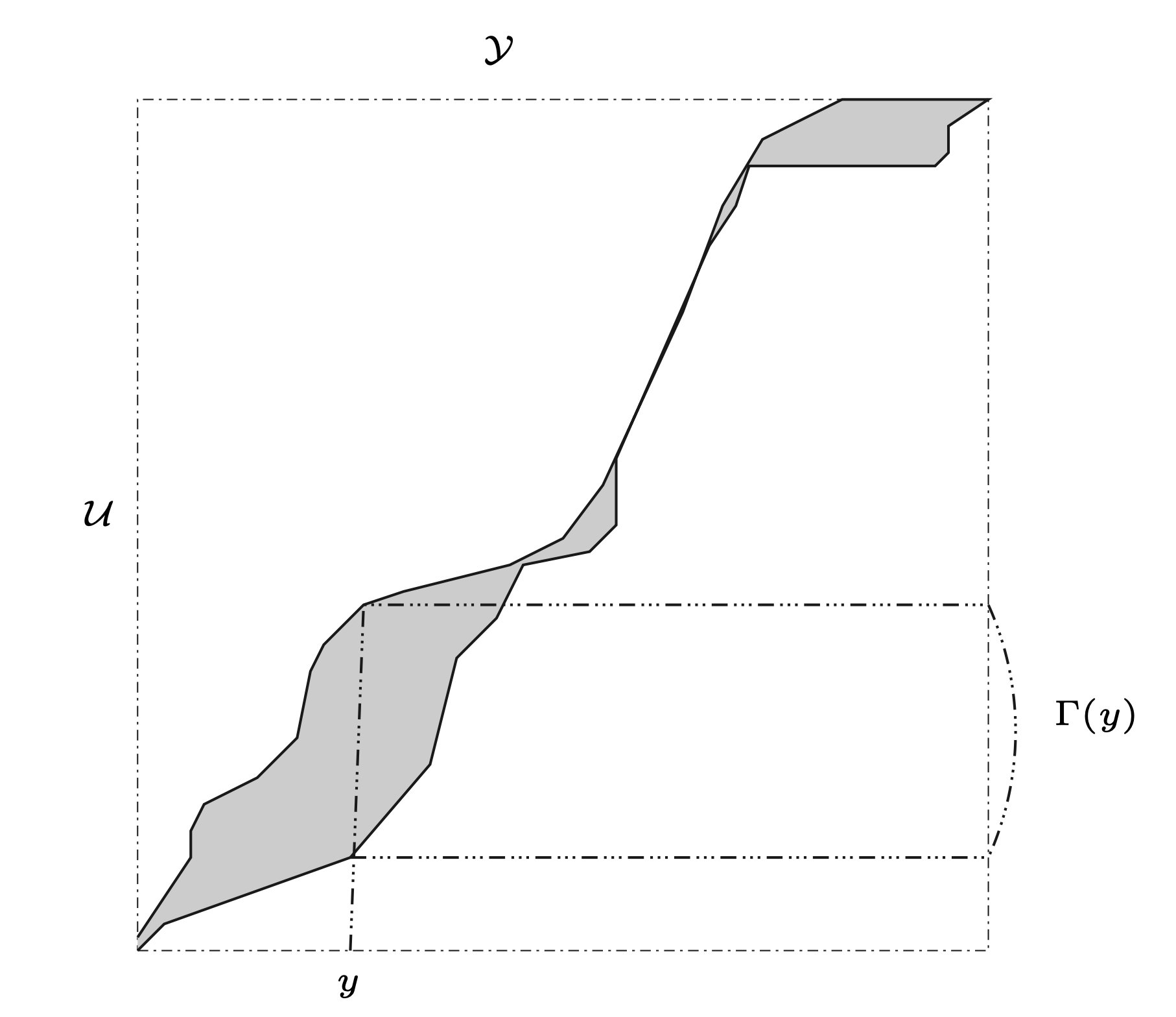}
\caption{Incomplete structure} \label{correspondence}
\end{center}
\end{figure}

If we call $P$ the true data-generating process for the
observables, we say that the complete structure is well specified
if $P(A)=\nu(\gamma(A))$ for all Borel sets $A$, which, by
Dynkin's lemma, is equivalent to $P(A)=\nu(\gamma(A))$ for all
cells $A$ of the form $(-\infty,y]$, $y\in\mathbb{R}$, which is
immediately seen to be equivalent to
\begin{eqnarray}\sup_{A\in{\cal S}}\;(P(A)-\nu(\gamma(A)))=0
\label{dual Monge-Kantorovich with sets in complete
model}\end{eqnarray} where ${\cal
C}=\{(-\infty,y_1],(y_2,\infty):\;(y_1,y_2)\in\mathbb{R}^2\}$.\vskip4pt
(\ref{dual Monge-Kantorovich with sets in complete model}) is a
programming problem, and it will turn out to be very fruitful to
consider its Monge-Kantorovich dual formulation
\begin{eqnarray}\inf_{\pi\in{\cal M}(P,\nu)}\int_{\mathbb{R}^2}1_{\{u\ne\gamma(y)\}}\;
\pi(dy,du)=0,\label{primal Monge-Kantorovich with complete
model}\end{eqnarray} where $1_{\{x\in A\}}$ denotes the indicator
function of the set $A$, and the infimum is taken over all joint
probability measures with marginals $P$ and $\nu$. The latter is a
mass transportation (or ``generalized matching'') problem, where
mass is transported from the set of observables to the set of
unobservables with zero-one cost of transportation associated with
violations of the constraint $u=\gamma(y)$.\vskip4pt This
formulation can be interpreted as the existence of a probability
that is concentrated on the structure, or alternatively, to the
existence of a coupling
between the random variable $Y$
with law $P$ and the random variable $U$ with law $\nu$, i.e. the
existence of $\pi$ with marginals $P$ and $\nu$ such that
\begin{eqnarray} \pi(U\ne\gamma(Y))=0.\label{marginal problem with
complete model}\end{eqnarray} We shall show that this dual
representation of the hypothesis of correct specification has a
natural generalization to the case of incomplete
structures.\vskip4pt Turning to empirical versions of the problem,
we can consider the statistic obtained by replacing $P$ by the
empirical distribution $P_n$ of a sample of independent and
identically distributed variables with law $P$, we obtain
\begin{eqnarray}\inf_{\pi\in{\cal M}(P,\nu)}\int_{\mathbb{R}^2}1_{\{u\ne\gamma(y)\}}\;
\pi(dy,du),\label{empirical primal Monge-Kantorovich with complete
model}\end{eqnarray} where the infimum is taken over probabilities
$\pi$ with marginals $P_n$ and $\nu$. By the above mentioned
duality, the latter is equal to
\begin{eqnarray*}\sup_{A\in{\cal B}_{\cal Y}}\;(P_n(A)-\nu(\gamma(A))),
\end{eqnarray*} with ${\cal B}_{\cal Y}$ the class of Borel sets.\vskip4pt
The last step is to determine a class of sets that is small enough
to allow determination of the limiting behaviour of the statistic,
i.e. we need to class of sets to be $P$-Donsker, and large enough
that the values of $\nu(\gamma(.))$ over all Borel sets are
determined by the latter's values on the restricted class. The
class ${\cal C}$ satisfies both requirements, and the resulting
test statistic is
\begin{eqnarray}\sup_{A\in{\cal C}}\;(P_n(A)-\nu(\gamma(A)))=\sup_{y\in\mathbb{R}}
|P_n(-\infty,y]-\nu(\gamma(-\infty,y])|, \label{Kolmogorov-Smirnov
with complete model}\end{eqnarray} which is exactly the
Kolmogorov-Smirnov specification test statistic.\vskip4pt We shall
essentially follow these same steps to show equivalence between
formulations of the hypothesis of correct specification and to
derive a test of specification when the bijection $\gamma$ is
replaced by a correspondence $\Gamma$, as in
figure~\ref{correspondence}. Then we shall consider parameterized
versions of the structure where both $\Gamma$ and $\nu$ depend on
a parameter $\theta$, and form confidence regions with all values
of $\theta$ such that the specification of model
$(\Gamma_{\theta},\nu_{\theta})$ is not rejected.\vskip6pt

\vskip6pt

\section{Incomplete model specifications}
We consider a very general econometric specification, thereby
posing the problem exactly as in \cite{Jovanovic:89} which
was an inspiration for this work. Variables under consideration
are divided into two groups.
\begin{itemize}\item Latent variables, $u\in {\cal U}$.
The vector $u$ is not observed by the analyst, but some of its
components may be observed by the economic actors. ${\cal U}$ is a
complete, metrizable and separable topological space (i.e. a
Polish space). \item Observable variables, $y\in {\cal
Y}=\mathbb{R}^{d_y}$. The vector $y$ is observed by the analyst.

\end{itemize} The Borel sigma-algebras of ${\cal Y}$ and ${\cal U}$ will be
respectively denoted ${\cal B}_{\cal Y}$ and ${\cal B}_{\cal U}$.
Call $P$ the Borel probability measure that represents the true
data generating process for the observable variables, and $\nu$
the hypothesized data generating processes for the latent
variables. The structure is given by a relation between observable
and latent variables, i.e. a subset of ${\cal Y}\times{\cal U}$,
which we shall write as a multi-valued mapping from ${\cal Y}$ to
${\cal U}$ denoted by $\Gamma$. Finally, the set of Borel
probability measures on $({\cal Y}\times{\cal U},\sigma({\cal
B}_{\cal Y}\times{\cal B}_{\cal U}))$ with marginals $P$ and $\nu$
is denoted by ${\cal M}(P,\nu)$. Whenever there is no ambiguity,
we shall adopt the de Finetti notation $\mu f$ to denote the
integral of $f$ with respect to $\mu$.\vskip4pt

\subsection{Examples}
\vskip8pt\noindent {\bf Example 1: Sample selection and other
models with missing counterfactuals.} The typical Heckman sample
selection models require very strong and often implausible
assumptions to guarantee identification. Weaker assumptions, such
as certain forms of monotonicity are plausible and restrict
significantly the identified set without reducing it to a
singleton. As an illustration of our formulation in this case,
consider for instance the classical set-up in
\cite{HV:2001}. We observe $(Y,D,W)$, where $Y$ is the
outcome variable, $D$ is an indicator variable for the receipt of
treatment, and $Z$ is a vector of instruments (we implicitly
condition the model on exogenous observable covariates). The
outcome variable is generated as follows:
\begin{eqnarray*}Y=DY_1+(1-D)Y_0,\end{eqnarray*} where $Y_0$
is the binary potential outcome if the individual does not receive
treatment, and $Y_1$ is the binary potential outcome if the
individual does receive treatment. The model is completed with the
specification of $D$ as follows:
\begin{eqnarray*}D=1_{\{g(Z)\geq U\}},\end{eqnarray*} where $g$ is a
measurable function and $U$ is uniformly distributed on $[0,1]$
(without loss of generality). The model can be written in the form
of a multi-valued mapping $\Gamma$ from observable to
unobservables in the following way:
\begin{eqnarray*}(y,d,z)&\longmapsto&\{(u,y_1,y_0)\in\Gamma(y,d,z)\}\\
(1,1,z)&\longmapsto&[\,0,g(z)]\times\{1\}\times\{0,1\}\\
(1,0,z)&\longmapsto&(g(z),1]\times\{0,1\}\times\{1\}\\
(0,1,z)&\longmapsto&[\,0,g(z)]\times\{0\}\times\{0,1\}\\
(0,0,z)&\longmapsto&(g(z),1]\times\{0,1\}\times\{0\}\end{eqnarray*}\vskip6pt

\vskip8pt\noindent {\bf Example 2: Returns to schooling.} Consider
a general specification for the returns to education, where income
$Y$ is a function of years of education $E$, other observable
characteristics $X$ and unobserved ability $U$ as $Y=G(E,X,U)$.
$G$ can be inverted as a multi-valued mapping to yield a
correspondence $U=\Gamma(Y,E,X)$.\vskip4pt

\vskip8pt\noindent {\bf Example 3: Censored data structures.}
Models with top-censoring or positive censoring such as Tobit
models fall in this class. A classic problem where identification
fails is regression with interval censored outcomes: the
observables variables are the pairs $(Y_*,Y^*,X)$ of upper and
lower values for the dependent variable, and the explanatory
variables. The correspondence describing the structure is
\begin{eqnarray*}\Gamma_{\theta}(y_*,y^*,x)=[y_*-x'\theta,y^*+x'\theta].\end{eqnarray*}

\vskip8pt\noindent {\bf Example 4: Games with multiple
equilibria.} Very large classes of economic models become
estimable with this approach, when one allows the object of
interest to be the identified set of parameters as opposed to
single parameter values. A simple class of examples is that of
models defined by a set of Nash rationality constraints. Suppose
the payoff function for player $j$, $j=1,\ldots,J$ is given by
\begin{eqnarray*}\Pi_j(S_j,S_{-j},X_j,U_j;\theta),\end{eqnarray*} where
$S_j$ is player $j$'s strategy and $S_{-j}$ is their opponents'
strategies. $X_j$ is a vector of observable characteristics of
player $j$ and $U_j$ a vector of unobservable determinants of the
payoff. Finally $\theta$ is a vector of parameters. Pure strategy
Nash equilibrium conditions
\begin{eqnarray*}\Pi_j(S_j,S_{-j},X_j,U_j;\theta)\geq
\Pi_j(S,S_{-j},X_j,U_j;\theta),\;\mbox{for all}\;S\end{eqnarray*}
define a correspondence $\Gamma_{\theta}$ from unobservable player
characteristics to observable variables $(S,X)$.\vskip6pt

\vskip8pt\noindent {\bf Example 5: Entry models.} Consider the
special case of example~4 proposed by \cite{Jovanovic:89}.
The payoff functions are
\begin{eqnarray*}\Pi_1(x_1,x_2,u)=(\lambda x_2-u)1_{\{x_1=1\}},\\
\Pi_2(x_1,x_2,u)=(\lambda x_1-u)1_{\{x_2=1\}},\end{eqnarray*}
where $x_i\in\{0,1\}$ is firm i's action, and $u$ is an exogenous
cost. The firms know their cost; the analyst, however, knows only
that $u\in[0,1]$, and that the structural parameter $\lambda$ is
in $(0,1]$. There are two pure strategy Nash equilibria. The first
is $x_1=x_2=0$ for all $u\in[0,1]$. The second is $x_1=x_2=1$ for
all $u\in[0,\lambda]$ and zero otherwise. Since the two firms'
actions are perfectly correlated, we shall denote them by a single
binary variable $y=x_1=x_2$. Hence the structure is described by
the multi-valued mapping: $\Gamma(1)=[0,\lambda]$ and
$\Gamma(0)=[0,1]$. In this case, since $y$ is Bernoulli, we can
write $P=(1-p,p)$ with $p$ the probability of a 1. For the
distribution of $u$, we consider a parametric exponential family
on $[0,1]$.\vskip6pt




We now turn to the definition of the null hypothesis of correct
specification and its empirical counterparts (in section~2), the
analysis of the properties of the test statistic (in section~3)
and the implementation and applications of the test (in
section~4).

\subsection{Null hypothesis of correct specification}
We wish to develop a procedure to detect whether the structure
$(\Gamma,\nu)$ and the distribution of observables are compatible.
First we explain what we mean by {\em compatible}. We start by
taking $P$, $\Gamma$ and $\nu$ as given and by considering three
natural formalizations of compatibility, a first representation
based on measurable selections of $\Gamma$, the second based on
the existence of a suitable probability measure with marginals $P$
and $\nu$ and a third based on Dempster's notion of maximal
plausibility.

\subsubsection{Equilibrium selections}
It is very easily understood in the simple case where the link
$\Gamma$ between latent and observable variables is parametric and
$\Gamma=\gamma$ is measurable and single valued. Defining the
image measure of $P$ by $\gamma$ by
\begin{eqnarray}P\gamma^{-1}(A)=P\{y\in {\cal Y}|\;
\gamma(y)\in A\},\label{Probability image}\end{eqnarray} for all
$A\in {\cal B}_{\cal U}$, we say that the structure is well
specified if and only if $\nu=P\gamma^{-1}$. In the general case
considered here, $\Gamma$ may not be single valued, and its images
may not even be disjoint (which would be the case if it was the
inverse image of a single valued mapping from ${\cal U}$ to ${\cal
Y}$, i.e. a traditional function from latent to observable
variables). However, under a measurability assumption on $\Gamma$,
we can construct an analogue of the image measure, which will now
be a set $\mbox{Core}(\Gamma,P)$ of Borel probability measures on
${\cal U}$ (defined by (\ref{Core})), and the hypothesis of {\em
compatibility} of the restrictions on latent variable
distributions and on the structures linking latent and observable
variables will naturally take the form
\begin{eqnarray}\mbox{H}_0: \nu\in\mbox{Core}(\Gamma,P).\label{hypothesis}\end{eqnarray}

\vskip8pt\noindent {\bf Assumption 1:} $\Gamma$ has non-empty and
closed values, and for each open set ${\cal O}\subseteq {\cal U}$,
$\;\Gamma^{-1}({\cal O})=\{y\in {\cal Y}\;|\; \Gamma(y)\cap{\cal
O}\neq\varnothing\}\in{\cal B}_{\cal Y}$.\vskip6pt

\begin{figure}[htbp]
\begin{center}
\includegraphics[width=8cm]{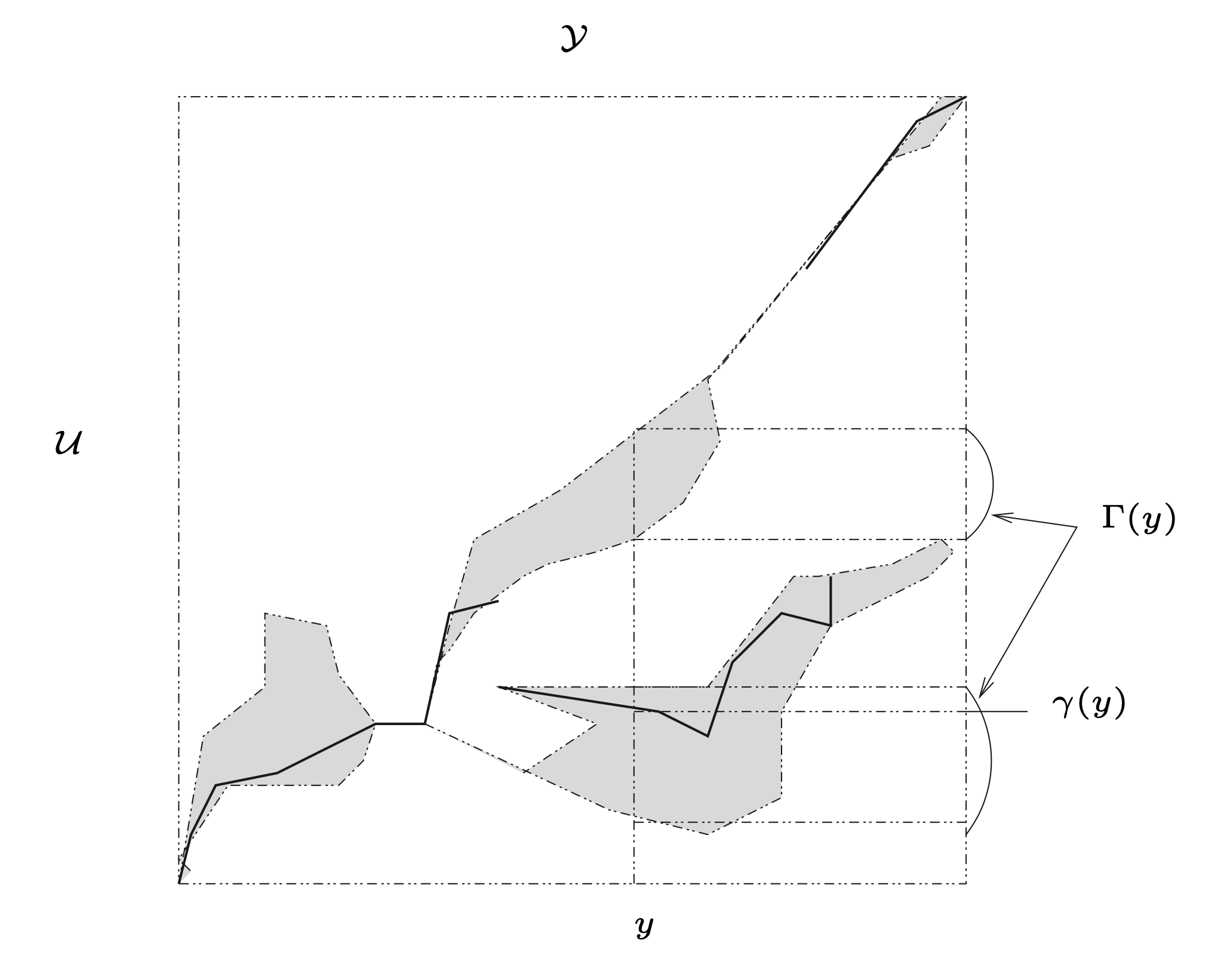}
\caption{Selection of a correspondence} \label{selection}
\end{center}
\end{figure}

To relate the present case to the intuition of the single-valued
case, it is useful to think in terms of single-valued {\em
selections} of the multi-valued mapping $\Gamma$, as in
figure~\ref{selection}.\vskip6pt

\vskip6pt

A measurable selection $\gamma$ of $\Gamma$ is a measurable
function such that $\gamma(y)\in\Gamma(y)$ for all $y\in {\cal
Y}$. The set of measurable selections of a multi-valued mapping
$\Gamma$ that satisfies Assumption~1 is denoted Sel($\Gamma$)
(which is known to be non-empty by the
Rokhlin-Kuratowsky-Ryll-Nardzewski Theorem).
To each selection $\gamma$ of $\Gamma$, we can
associate the image measure of $P$, denoted $P\gamma^{-1}$,
defined as in (\ref{Probability image}).\vskip6pt

It would be tempting to reformulate the compatibility condition as
the requirement that at least one selection $\gamma$ in
Sel($\Gamma$) is such that $\nu=P\gamma^{-1}$. However, such a
requirement implies that $\gamma$ corresponds to the equilibrium
that is always selected. Under such a requirement, if for a given
observable value the structure does not specify which value of the
latent variables gave rise to it, the latter is nonetheless fixed.
Hence two identical observed realizations in the sample of
observations necessarily arose from the same realization of the
latent variables. We argue, however, that if the structure does
not specify an equilibrium selection mechanism, there is no reason
to assume that each observation is drawn from the same
equilibrium.\vskip6pt


Allowing endogenous equilibrium selection of unknown form is
equivalent to allowing the existence of an arbitrary distribution
on the set of $P\gamma^{-1}$ when $\gamma$ spans Sel$(\Gamma)$ (as
opposed to a mass on one particular $P\gamma^{-1}$). A Bayesian
formulation of the problem would entail a specification of this
distribution. Here, we stick to the given specification in leaving
it completely unspecified.
\vskip6pt

Hence, we argue that the correct reformulation of the
compatibility condition is that $\nu$ can be written as a mixture
of probability measures of the form $P\gamma^{-1}$, where $\gamma$
ranges over Sel($\Gamma$). However, as the following example show,
even for the simplest multi-valued mapping, the set of measurable
selections is very rich, let alone the set of their mixtures.

\vskip8pt\noindent{\bf Example:} Consider the multi-valued mapping
\begin{eqnarray*}\Gamma:\;[0,1]\rightrightarrows[0,1]\end{eqnarray*}
defined by $\Gamma(x)=\{0,x\}$ for all $x$. The collection of
measurable selections of $\Gamma$ is indexed by the class of Borel
subsets of $[0,1]$. Indeed, a representative measurable selection
of $\Gamma$ is $\gamma_B$, such that $\gamma_B(x)=x1_{\{x\in B\}}$
for any Borel subset $B$ of $[0,1]$, where $1_{\{x\in B\}}$
denotes the indicator function which equals one when $x\in B$ and
zero otherwise.\vskip4pt

Hence, it will be imperative to give manageable equivalent
representations of such a mixture, as is done in Theorem~1 below.

\subsubsection{Existence of a suitable joint probability}
The second natural representation of compatibility of the
distribution $P$ of observables and the structure $(\Gamma,\nu)$
is based on the existence of probability measures on the product
${\cal Y}\times{\cal U}$ that admit $P$ and $\nu$ as
marginals.\vskip4pt In the benchmark case of $\Gamma=\gamma$
one-to-one, the structure imposes a stringent constraint on pairs
$(y,u)$, namely that $u=\gamma(y)$. So the admissible region of
the product space is the graph of $\gamma$, i.e. the set
\begin{eqnarray*}\mbox{Graph}\;\gamma=\{(y,u)\in{\cal Y}\times{\cal U}:\;
u=\gamma(y)\}.\end{eqnarray*} The compatibility condition
described above, namely $P\gamma^{-1}=\nu$ is equivalent to the
existence of a probability measure on the product space that is
supported by Graph~$\gamma$ (i.e. that gives probability zero
outside the constrained region defined by the structure) and
admits $P$ and $\nu$ as marginals.\vskip4pt This generalizes
immediately to the case of $\Gamma$ multi-valued, as the existence
of a probability measure that admits $P$ and $\nu$ as marginals,
and that is supported on the constrained region
\begin{eqnarray}\mbox{Graph}\;\Gamma=\{(y,u)\in{\cal Y}\times{\cal U}:\;
u\in\Gamma(y)\},\label{graph}\end{eqnarray} in other words, a
probability measure that admits $P$ and $\nu$ as marginals and
gives probability zero to the event $U\notin\Gamma(Y)$, where $U$
and $Y$ are random elements with probability law $\nu$ and $P$
respectively (namely (\ref{marginal problem}) below).

\subsubsection{Dempster plausibility} \cite{Dempster:67} suggests to
consider the smallest reliability that can be associated with the
event $B\in{\cal B}_{\cal U}$ as the {\em belief function}
\begin{eqnarray*}\underline{P}(A)=P\{y\in{\cal Y}\;|
\;\Gamma(y)\subseteq B\}\end{eqnarray*} and the largest
plausibility that can be associated with the event $B$ as the {\em
plausibility function}
\begin{eqnarray*}\overline{P}(A)=P\{y\in{\cal Y}\;|
\;\Gamma(y)\cap B\neq\varnothing\}\end{eqnarray*} the two being
linked by the relation
\begin{eqnarray}\overline{P}(A)=1-\underline{P}(A^c),\label{conjugates}\end{eqnarray}
which prompted some authors to call them {\em conjugates} or {\em
dual} of each other.\vskip4pt

\begin{figure}[htbp]
\begin{center}
\includegraphics[width=8cm]{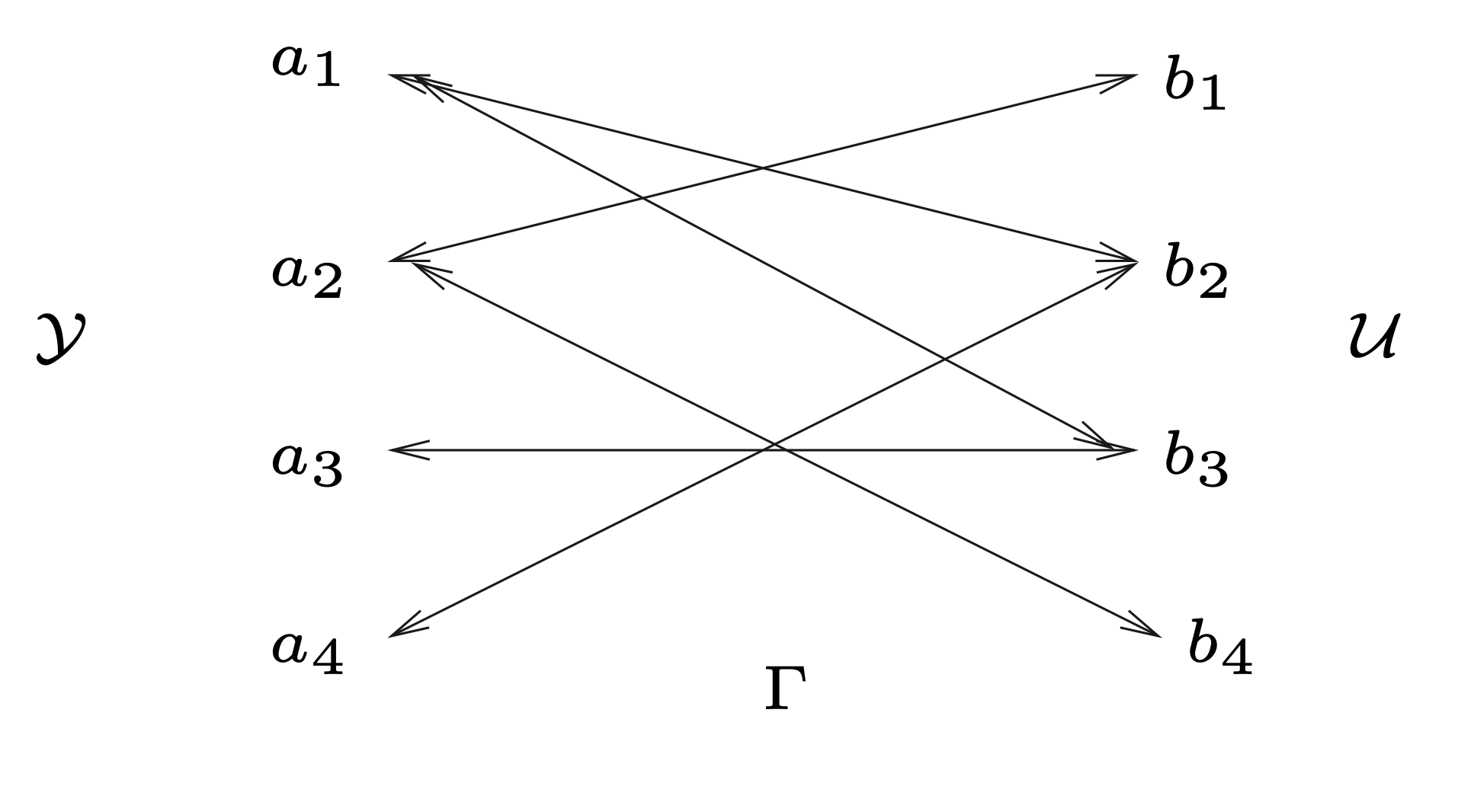}
\caption{Graph of the correspondence $\Gamma$ in a finite case.
The event $\{a_3\}$ always gives rise to the event $\{b_3,b_4\}$,
whereas event $\{a_4\}$ never does, so it is natural to constrain
the probability of the event $\{b_3,b_4\}$ by the upper bound
$P(\{a_1,a_2,a_3\})$ and the lower bound $P(\{a_3\})$.}
\label{dempster}
\end{center}
\end{figure}

A natural way to construct a set of probability measures is to
consider all probability measures that do not exceed the largest
plausibility that can be associated with a set, and that, as a
result of (\ref{conjugates}), are larger than the smallest
reliability associated with a set. We thus form the {\em core} of
the belief function\footnote{The name Core is standard in the
literature to denote the set of probability measures satisfying
(\ref{plausibility constraint}). It seems to originate from D.
Gillies' 1953 Princeton PhD thesis on ``some theorems on n-person
games.'' For finite sets, the core is non-empty by the
Bondareva-Shapley theorem. In the present more general context,
the non-emptiness of the core will follow from the equivalence of
(i) and (iv) of Theorem~1 below, and the existence of measurable
selections of $\Gamma$ under assumption~1.}:
\begin{eqnarray}\mbox{Core}(\Gamma,P)&=&\{
\mu\in\Delta({\cal U})\;|\;\forall
B\in{\cal B}_{\cal U},\,\mu(B)\geq\underline{P}(B)\}\label{Core}\\
&=&\{ \mu\in\Delta({\cal U})\;|\;\forall B\in{\cal B}_{\cal
U},\,\mu(B)\leq\overline{P}(B)\}\nonumber\end{eqnarray} where the
first equality can be taken as a definition, and the second
follows immediately from (\ref{conjugates}). It is well known that
$\mbox{Core}(\Gamma,P)$ is non-empty, and another natural
representation of the compatibility of the distribution $P$ of
observables with the structure $(\Gamma,\nu)$ is that $\nu$
belongs to $\mbox{Core}(\Gamma,P)$, in other words, that $\nu$
satisfies $\nu(B) \leq P(\{y\in{\cal Y}:\; \Gamma(y)\cap
B\ne\varnothing\})$ for all $B\in{\cal B}_{\cal U}$.
Figure~\ref{dempster} illustrates this requirement in the case of
finite sets. \vskip6pt

\subsubsection{Equivalence of compatibility representations}
The following theorem shows that the three representations
discussed above are, in fact, equivalent. In addition, two more
equivalent formulations are presented that will be used in the
empirical formulations in the next section.

\vskip8pt\noindent{\bf Theorem 1:} Under assumption~1, the
following statements are equivalent:\begin{itemize} \item[(i)]
$\nu$ is a mixture of images of $P$ by measurable selections of
$\Gamma$, (i.e. $\nu$ is in the weak closed convex hull of
$\{P\gamma^{-1};\;\gamma\in{\mathrm{Sel}}(\Gamma)\}$).\item[(ii)]
There exists for $P$-almost all $y\in {\cal Y}$ a probability
measure $\pi_{\nu}(y,.)$ on ${\cal U}$ with support $\Gamma(y)$,
such that
\begin{eqnarray}\nu(B)=\int_{\cal Y}\pi_{\nu}(y,B)\;P(dy),\;\mbox{all}\;
B\in{\cal B}_{\cal
U}.\label{disintegration}\end{eqnarray}\item[(iii)] If $U$ and $Y$
are random elements with respective distributions $P$ and $\nu$,
there exists a probability measure $\pi\in{\cal M}(P,\nu)$ that is
supported on the admissible region, i.e. such that
\begin{eqnarray}\pi(U\notin\Gamma(Y))=0.\label{marginal problem}\end{eqnarray}
\item[(iv)] The probability assigned by $\nu$ to an event in
$B\in{\cal B}_{\cal U}$ is no greater than the largest
plausibility associated with $B$ given $P$ and $\Gamma$, i.e.
\begin{eqnarray}\nu(B) \leq P(\{y\in{\cal Y}:\;
\Gamma(y)\cap B\ne\varnothing\})\label{plausibility
constraint}\end{eqnarray} \item[(v)] For all $A\in{\cal B}_{\cal
Y}$, we have
\begin{eqnarray}P(A)\leq\nu(\Gamma(A)).\label{Kolmogorov-Smirnov Representation}\end{eqnarray}
\end{itemize}\vskip6pt

\noindent{\bf Remark 1:} The weak topology on $\Delta({\cal U})$,
the set of probability measures on ${\cal U}$, is the topology of
convergence in distribution. $\Delta({\cal U})$ is also Polish,
and the weak closed convex hull of
$\{P\gamma^{-1};\;\gamma\in{\mathrm{Sel}}(\Gamma)\}$ is indeed the
collection of arbitrary mixtures of elements of
$\{P\gamma^{-1};\;\gamma\in{\mathrm{Sel}}(\Gamma)\}$.

\vskip8pt\noindent{\bf Remark 2:}
Notice that (\ref{disintegration}) looks like a disintegration of
$\nu$,
and indeed, when $\Gamma$ is the inverse image of a single-valued
measurable function (i.e. when the structure is given by a
single-valued measurable function from latent to observable
variables), the probability kernel $\pi_{\nu}$  is exactly the
$(P,\Gamma^{-1})$-disintegration of $\nu$, in other words,
$\pi_{\nu}(y,.)$ is the conditional probability measure on ${\cal
U}$ under the condition $\Gamma^{-1}(u)=\{y\}$. Hence
(\ref{disintegration}) has the interpretation that a random
element with distribution $\nu$ can be generated as a draw from
$\pi_{\nu}(y,.)$ where $y$ is a realization of a random element
with distribution $P$.\vskip6pt

\vskip8pt\noindent{\bf Remark 3:} As will be explained later, our
test statistic will be based on violations of representation (v),
which is the dual formulation of (iii) seen as a Monge-Kantorovich
optimal mass transportation solution.\vskip6pt

\vskip8pt\noindent{\bf Remark 4:} Equivalence of (i) and (iii) is
a generalization of proposition~1 of \cite{Jovanovic:89} to
the case where $P$ is not necessarily atomless and ${\cal U}$ not
necessarily compact. Notice that relative to
\cite{Jovanovic:89}, the roles of ${\cal Y}$ and ${\cal U}$
are reversed for the purposes of specification testing. As
discussed in the second remark following proposition~1 mentioned
above, atomlessness of the distribution of latent variables is
innocuous as long as ${\cal U}$ is rich enough. However,
atomlessness of the distribution of observables isn't innocuous,
since it rules out many of the relevant applications.\vskip6pt

Note that since as a multivalued function, $\Gamma$ is always
invertible, and Assumption~1 holds for $\Gamma$ if and only if it
holds for $\Gamma^{-1}$, 
the roles of $P$ and $\nu$ can be
interchanged in the formulations. In some cases, the symmetric
formulation, with the roles of $P$ and $\nu$ interchanged, is
useful, so we state it for completeness below:

\vskip8pt\noindent{\bf Theorem 1':} Under assumption~1, the
following statements are equivalent, and are also equivalent to
each of the statements in Theorem~1:\begin{itemize} \item[(i')]
$P$ is a mixture of images of $\nu$ by measurable selections of
$\Gamma^{-1}$, (i.e. $P$ is in the weak closed convex hull of
$\{\nu\gamma^{-1};\;\gamma\in{\mathrm{Sel}}(\Gamma^{-1})\}$).\item[(ii')]
There exists for $\nu$-almost all $u\in {\cal U}$ a probability
measure $\pi_{P}(u,.)$ on ${\cal Y}$ with support
$\Gamma^{-1}(u)$, such that
\begin{eqnarray}P(A)=\int_{\cal U}\pi_{P}(u,A)\;\nu(du),\;\mbox{all}\;
A\in{\cal B}_{\cal
Y}.\label{disintegration2}\end{eqnarray}\item[(iii')] is identical
to Theorem~1(iii). \item[(iv')] The probability assigned by $P$ to
an event in $A\in{\cal B}_{\cal Y}$ is no greater than the largest
plausibility associated with $A$ given $\nu$ and $\Gamma^{-1}$,
i.e.
\begin{eqnarray}P(A) \leq \nu(\{u\in{\cal U}:\;
\Gamma^{-1}(u)\cap A\ne\varnothing\})\label{plausibility
constraint2}\end{eqnarray} \item[(v')] For all $B\in{\cal B}_{\cal
U}$, we have
\begin{eqnarray}\nu(B)\leq P(\Gamma^{-1}(B)).\label{Kolmogorov-Smirnov Representation2}\end{eqnarray}
\end{itemize}\vskip6pt

\vskip8pt\noindent{\bf Remark 1:} The reason for giving this
second theorem is that some of the new formulations will more
amenable to forming empirical counterparts. \vskip6pt

\section{Empirical formulations} Each of the theoretical
formulations of correct specification of the structure given in
Theorems~1 and~1' has empirical counterparts, obtained essentially
by replacing $P$ by an estimate such as $P_n$ in the formulations.
The equivalence of the theoretical formulations does not
necessarily entail equivalence of the empirical counterparts,
especially in the cases where they rely on a choice of distance on
the (metrizable) space of probability measures on $({\cal Y},{\cal
B}_{\cal Y})$ or $({\cal U},{\cal B}_{\cal U})$. Hence we need to
consider the relations existing between the different empirical
counterparts. We shall form our test statistic based on the
empirical formulation relative to (v), so the reader may jump to
section~2.4 without loss of continuity.

\subsection{Empirical representations relative to (i)}
For this empirical formulation, we consider (i') from Theorem~1'.
We denote Core$(\Gamma^{-1},\nu)$ the set of arbitrary mixtures of
$\nu\gamma^{-1}$ when $\gamma$ spans Sel$(\Gamma^{-1})$, and
denoting by $d$ a choice of metric on the space of probability
measures on $({\cal Y}, {\cal B}_{\cal Y})$, the null can be
reformulated as
\begin{eqnarray*}d(P,\mbox{Core}(\Gamma^{-1},\nu)):=\inf_{\mu\in
{\mathrm{Core}}(\Gamma^{-1},\nu)}d(P,\mu)=0.\end{eqnarray*} Hence
the empirical version is obtained by replacing $P$ by an estimate
such as $P_n$ to yield
\begin{eqnarray*}d(P_n,\mbox{Core}(\Gamma^{-1},\nu)).
\label{empirical formulation (i)}\end{eqnarray*} It will naturally
depend on the specific choice of metric.\vskip4pt To see the
relation between this and other empirical formulations, consider
the Kolmogorov-Smirnov metric defined by
\begin{eqnarray*}d_{{\mathrm KS}}(\mu_1,\mu_2)=\sup_{A\in{\cal B}_{\cal
Y}}(\mu_1(A)-\mu_2(A))\end{eqnarray*} for any two probability
measures $\mu_1$ and $\mu_2$ on $({\cal Y}, {\cal B}_{\cal Y})$.
With this choice of metric, we can derive conditions under which
the equalities
\begin{eqnarray*}d_{{\mathrm
KS}}(P_n,\mbox{Core}(\Gamma^{-1},\nu))&=&\inf_{\gamma\in{\mathrm
Sel}(\Gamma^{-1})}\sup_{A\in{\cal B}_{\cal Y}}
(P_n(A)-\nu\gamma^{-1}(A))\\&=&\sup_{A\in{\cal B}_{\cal Y}}
\inf_{\gamma\in{\mathrm
Sel}(\Gamma)}(P_n(A)-\nu\gamma(A))\\&=&\sup_{A\in{\cal B}_{\cal
Y}}(P_n(A)-\nu(\Gamma(A)))\end{eqnarray*} hold, and therefore this
empirical formulation is equivalent to empirical formulations
based on (iii), (iv), and (v) below.

\subsection{Empirical representations relative to (ii)}
We consider (ii) from Theorem~1 and $d$ a metric on the space of
probability measures on $({\cal U}, {\cal B}_{\cal U})$. Under the
null hypothesis, let $\pi_{\nu}$ be the family of kernels defined
in (ii) of Theorem~1. Denoting $\mu f$ the integral of a function
$f$ by a measure $\mu$, we can write (ii) as
$d(\nu,P\pi_{\nu})=0$, which admits $d(\nu,P_n\pi_{\nu})$ as
empirical counterpart, and the latter is equal to
$d(P\pi_{\nu},P_n\pi_{\nu})$. A notable aspect of this empirical
formulation is that for many choices of metric $d$ or indeed
pseudo-metric (such as relative entropy), it will take the form of
a functional of the empirical process
$\mathbb{G}_n:=\sqrt{n}(P_n-P)$ applied to the functions
$y\mapsto\pi_{\nu}(y).$ Different Goodness-of-fit tests can
therefore be generalized within a single framework. The difficulty
here of course is that the kernel $\pi_{\nu}$ depends on the
unknown $P$ in a complicated way through the integral equation
(\ref{disintegration}).

\subsection{Empirical representation relative to (iii)}
In view of representation (iii) of Theorem~1, i.e. equation
(\ref{marginal problem}), the null can be reformulated as the
following Monge-Kantorovich mass transportation problem
\begin{eqnarray} \min_{\pi\in{\cal M}(P,\nu)}\;\int_{{\cal Y}\times{\cal U}}
1_{\{u\notin\Gamma(y)\}}\;\pi(dy,du)=0,\label{primal
Monge-Kantorovich}\end{eqnarray} where the transportation cost
function $1_{\{u\notin\Gamma(y)\}}$ is an indicator penalty for
violation of the structure.

We now consider the empirical version of this Monge-Kantorovich
problem, replacing $P$ by the empirical distribution $P_n$ to
yield the functional
\begin{eqnarray}T^*(P_n,\Gamma,\nu)=\min_{\pi\in{\cal M}(P_n,\nu)}\;\int_{{\cal Y}\times{\cal U}}
1_{\{u\notin\Gamma(y)\}}\;\pi(dy,du).\label{infeasible test
statistic}\end{eqnarray}

We shall see below that it is equal to the empirical formulations
relative to (iv) and (v).

\subsection{Empirical representation relative to (iv) and (v)}
Since formulations (iv) and (v) from Theorem~1 can be rewritten
\begin{eqnarray*}\sup_{A\in{\cal B}_{\cal Y}}(P(A)-\nu(\Gamma(A)))=0,\end{eqnarray*}
the following empirical formulation seems the most natural:
\begin{eqnarray*}\sup_{A\in{\cal B}_{\cal Y}}(P_n(A)-\nu(\Gamma(A))).\end{eqnarray*}

The following Theorem states the equivalence between the latter
and the empirical formulation derived from (iii):

\vskip8pt\noindent{\bf Theorem 2:} The following equalities hold:
\begin{eqnarray}T^*(P_n,\Gamma,\nu)&=&\max_{f\oplus g\leq\varphi}\;
\left(P_nf+\nu g\right)
\label{dual Monge-Kantorovich with functions}\\
&=&\sup_{A\in{\cal B}_{\cal
Y}}\;\left(P_n(A)-\nu(\Gamma(A))\right),\label{dual
Monge-Kantorovich with sets}\end{eqnarray} where
$\varphi(y,u)=1_{\{u\notin\Gamma(y)\}}$, and $f\oplus
g\leq\varphi$ signifies that the maximum in (\ref{dual
Monge-Kantorovich with functions}) is taken over all measureable
functions $f$ on ${\cal Y}$ and $g$ on ${\cal U}$ such that for
all $(y,u)$, $f(y)+g(u)\leq\varphi(y,u)$. \vskip6pt

We shall therefore take $T^*(P_n,\Gamma,\nu)$ as our starting
point to construct a test statistic in the following section.

\section{Specification test} We propose to adopt a test
statistic based on the dual Monge-Kantorovich formulation
(\ref{dual Monge-Kantorovich with sets}), in other words a
statistic that penalizes large values of (\ref{dual
Monge-Kantorovich with sets}). However, $T^*(P_n,\Gamma,\nu)$
seemingly involves checking condition (\ref{Kolmogorov-Smirnov
Representation}) on all sets in ${\cal B}_{\cal Y}$. We need to
elicit a reduced class of sets on which to check condition
(\ref{Kolmogorov-Smirnov Representation}). Call such a reduced
class ${\cal S}$, and the resulting statistic is
\begin{eqnarray}T_{\cal S}(P_n,\Gamma,\nu)=\sup_{A\in{\cal
S}}\;\left(P_n(A)-\nu(\Gamma(A))\right).\label{test
statistic}\end{eqnarray} ${\cal S}$ is the result of a formal
trade-off: it needs to be small enough to allow us to derive a
limiting distribution for a suitable re-scaling of
$T(P_n,\Gamma,\nu)$, and large enough to determine the direction
of the inequality $P-\nu\Gamma$, which corresponds to a
requirement that our test retain power against fixed
alternatives.\vskip6pt

To illustrate these requirements, we start by considering two
simple types of structures to be tested. First we shall consider
bijective structures (which correspond to our ``prelude''), then
the case where ${\cal Y}$ is finite.

\begin{itemize} \item {\bf Bijective structures:}
In the case where $\Gamma=\gamma$ is single-valued and bijective,
consider the following classes of cells in $\mathbb{R}^{d_{y}}$:
\begin{eqnarray*}{\cal
C}&=&\{(-\infty,y],(y,\infty):\;y\in\overline{\mathbb{R}}^{d_{y}}\}\\
\tilde{{\cal C}}&=&\{(-\infty,y]:\;y\in\mathbb{R}^{d_{y}}\}.
\end{eqnarray*} Notice that
\begin{eqnarray*}\sup_{A\in{\cal
C}}\;\left(P_n(A)-\nu(\gamma(A))\right)=\sup_{A\in\tilde{{\cal
C}}}\;\left|P_n(A)-\nu(\gamma(A))\right|\end{eqnarray*} and the
latter is the classical Kolmogorov-Smirnov specification test
statistic. Hence the choice of ${\cal C}$ for our reduced class
${\cal S}$ is suitable on both counts: we know, as was discussed
in the prelude, that ${\cal C}$ is a value-determining class for
probability measures, hence checking the inequality $P-\nu\gamma$
on the reduced class is equivalent to checking it on all
measurable sets. In addition, from Appendix~A1, we know that this
class is Vapnik-$\breve{\mbox{C}}$ervonenkis, and hence that
$\sqrt{n}T_{\cal C}(P_n,\gamma,\nu)=\sup_{A\in{\cal
C}}\mathbb{G}_n(A)$ converges weakly to the supremum of a
$P$-Brownian bridge, and the test of specification can be
constructed based on approximations of the quantiles through
simulations of the Brownian bridge or the bootstrap.

\item {\bf Discrete observables:} In the case where the
observables belong to a finite set, the power set $2^{\cal Y}$ is
finite, hence Vapnik-$\breve{\mbox{C}}$ervonenkis. This will be
sufficient to derive the limiting distribution of
$\sqrt{n}T_{2^{\cal Y}}(P_n,\Gamma,\nu)= \sqrt{n}\sup_{A\in2^{\cal
Y}}\;(P_n(A)-\nu(\Gamma(A)))$. Since class of whole subsets is
used, we do not need to worry about the competing requirements
that the class determine the direction of the inequality
$P-\nu\Gamma$.\end{itemize}

We shall consider the two requirements on the class of sets ${\cal
S}$ sequentially. First, in the next subsection, we derive the
asymptotic distribution of $T_{\cal S}(P_n,\Gamma,\nu)$ for a
given choice of ${\cal S}$. Then, in the following subsection, we
examine the power of the test based on $T_{\cal
S}(P_n,\Gamma,\nu)$, which amounts to linking the choice of the
class of sets ${\cal S}$ with classes of alternatives.

\subsection{Asymptotic analysis}
We start with a short heuristic description of the behaviour of
$T_{\cal S}(P_n,\Gamma,\nu)$ which will motivate some definitions
and constructions. We then give specific sets of conditions for
the asymptotic results to hold.

\subsubsection{Heuristic description of asymptotic behaviour}
\begin{figure}[htbp]
\begin{center}
\includegraphics[width=8cm]{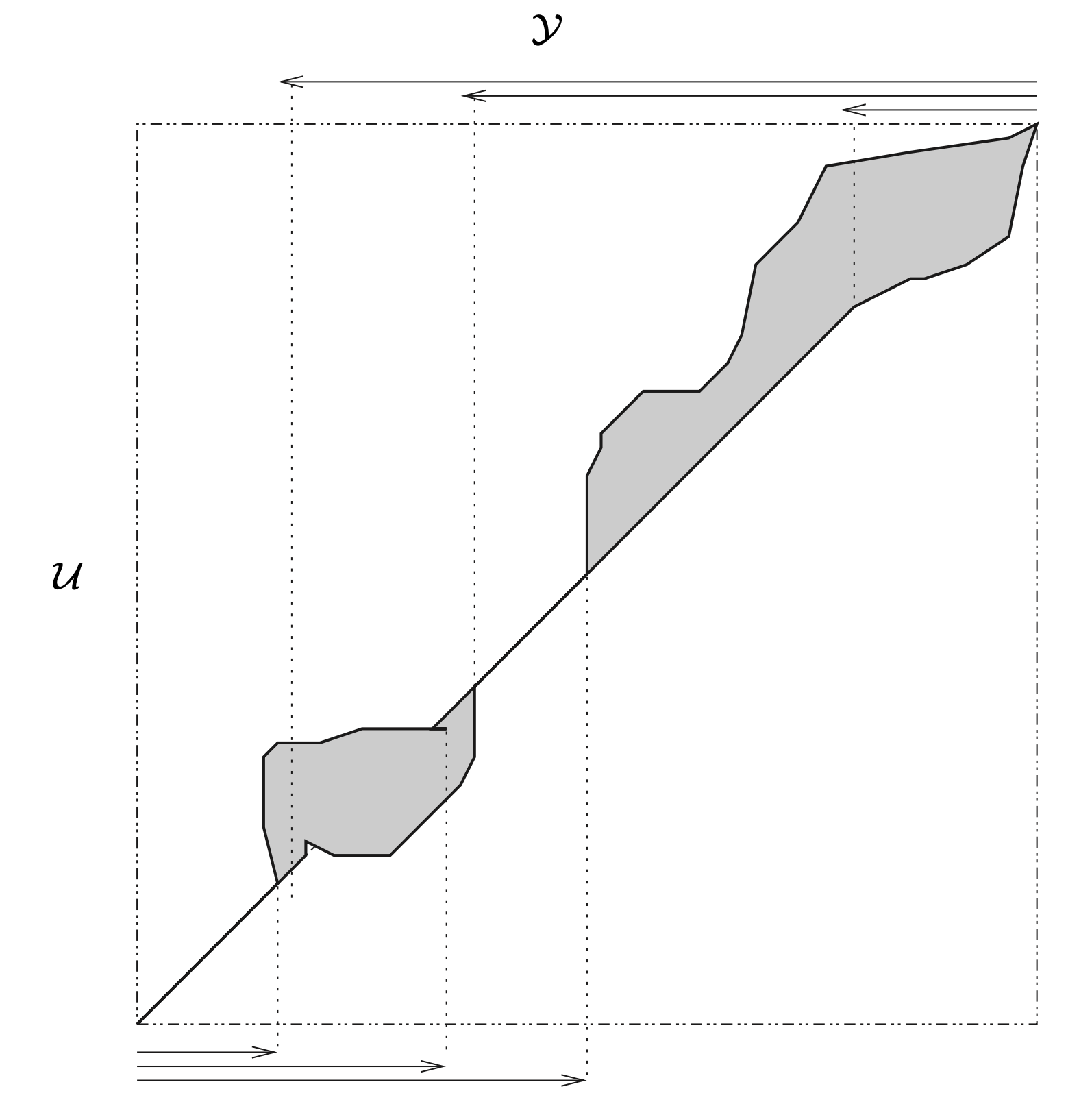}
\caption{Examples of sets in ${\cal C}_b$ (symbolized by the
arrows) in a correctly specified case ($P$ and $\nu$ are uniform,
hence correct specification corresponds to the graph of $\Gamma$
containing the diagonal).} \label{saturated}
\end{center}
\end{figure}

Under the null hypothesis H$_0$, we have
$P(A)-\nu(\Gamma(A))\leq0$ for all $A\in{\cal B}_{\cal Y}$.
Recalling that $\mathbb{G}_n$ is the empirical process
$\sqrt{n}(P_n-P)$, we have
\begin{eqnarray*}\sqrt{n}\,T_{\cal S}(P_n,\Gamma,\nu)&=&\sqrt{n}\sup_{A\in{\cal S}}
(P_n(A)-\nu(\Gamma(A)))\\&=&\sup_{A\in{\cal S}}
(\mathbb{G}_n(A)+\sqrt{n}(P(A)-\nu(\Gamma(A)))).\end{eqnarray*}
Unlike the case of the classical Kolmogorov-Smirnov test, the
second term in the previous display does not vanish under the
null, since the ``regions of indeterminacy'' allow
$\delta(A):=P(A)-\nu(\Gamma(A))$ to be strictly negative for some
sets $A\in{\cal S}$. What we know at this stage is that under the
null, we have
\begin{eqnarray*}\sqrt{n}\,T_{\cal
S}(P_n,\Gamma,\nu)=\sup_{A\in{\cal S}}
(\mathbb{G}_n(A)+\sqrt{n}(P(A)-\nu(\Gamma(A))))\leq
\sup_{A\in{\cal S}}\mathbb{G}_n(A),\end{eqnarray*} but relying on
this bound may lead to very conservative inference.\vskip6pt Note
that $\delta$ is independent of $n$, so that the scaling factor
$\sqrt{n}$ will pull the second term in the previous display to
$-\infty$ for all the sets where the inequality is strict. This
prompts the following definition, illustrated in
figure~\ref{saturated}:\vskip8pt

\vskip8pt\noindent{\bf Definition 3.1:} We denote the subclass of
sets from ${\cal S}$ where $P=\nu\Gamma$ by ${\cal S}_b$, i.e.
\begin{eqnarray*}{\cal S}_b:=\left\{A\in{\cal S}:\;
P(A)=\nu(\Gamma(A))\right\}.\end{eqnarray*}\vskip6pt

If the class ${\cal S}$ is a Vapnik-$\breve{\mbox{C}}$ervonenkis
class of sets, the empirical process converges weakly to the
$P$-Browninan bridge $\mathbb{G}$, i.e. a tight centered Gaussian
stochastic process with variance-covariance defined by
\begin{eqnarray*}\mathbb{E}\mathbb{G}(A_1)\mathbb{G}(A_2)=P(A_1\cap A_2)
-P(A_1)P(A_2),\end{eqnarray*} and the convergence is uniform over
the class ${\cal S}$ (i.e. the convergence is in $l^{\infty}({\cal
F})$, where ${\cal F}$ is the class of indicator functions of sets
in ${\cal S}$), so that by the continuous mapping theorem,
the supremum of the empirical process converges weakly to the
supremum of the Brownian bridge (for a detail of the proof, see
Appendix~A1).\vskip6pt

Under (mild) conditions that ensure that the function $\delta$
``takes off'' frankly from zero on ${\cal S}_b$ to negative values
on ${\cal S}\backslash{\cal S}_b$, the term $\sqrt{n}\,\delta$
dominates the oscillations of the empirical process, and the sets
in ${\cal S}\backslash{\cal S}_b$ drop out from the supremum in
the asymptotic expression, so that
\begin{eqnarray} \sqrt{n}\,T_{\cal
S}(P_n,\Gamma,\nu)\rightsquigarrow\sup_{A\in{\cal
S}_b}\;\mathbb{G}(A),\label{asymptotic distribution}\end{eqnarray}
where $\rightsquigarrow$ denotes weak convergence.
Naturally, since ${\cal S}_b$ depends on the unknown $P$, we need
to find a data dependent class of sets to approximate ${\cal
S}_b$. By the Law of Iterated Logarithm (see for instance page 476
of \cite{Dudley:2003}), we know that the empirical process
$\mathbb{G}_n$ is uniformly $O_p(\sqrt{\ln\ln n})$, so that if we
construct the data dependent class as in definition~2 below with a
bandwidth sequence $h=h_n>0$ satisfying
\begin{eqnarray}h_n+h_n^{-1}\sqrt{\frac{\ln\ln n}{n}}\rightarrow0,
\label{bandwidth}\end{eqnarray}
we shall pick out the sets in ${\cal S}_b$ asymptotically.

\vskip8pt\noindent{\bf Definition 3.2:} We denote the data
dependent subclass of sets from ${\cal S}$ where
$P_n\geq\nu\Gamma-h$ by $\hat{\cal S}_{b,h}$, i.e.
\begin{eqnarray*}\hat{\cal S}_{b,h}:=\left\{A\in{\cal S}:\;
P_n(A)\geq\nu(\Gamma(A))-h\right\}.\end{eqnarray*}

This data dependent class of sets allows us to approximate the
distribution of $T_{\cal S}(P_n,\Gamma,\nu)$ based on the
following limiting result
\begin{eqnarray}\sup_{A\in\hat{\cal S}_{b,h_n}}\mathbb{G}(A)
\rightsquigarrow\sup_{A\in{\cal
S}_b}\mathbb{G}(A)\label{asymptotic approximation}\end{eqnarray}
under requirement (\ref{bandwidth}) on the bandwidth sequence
$h_n$, and the additional requirement that
\begin{eqnarray}h_n (\ln\ln n)\rightarrow0,\label{bandwidth2}\end{eqnarray}
which allows to control local oscillations of the empirical
process as well. Note that (\ref{bandwidth}) and
(\ref{bandwidth2}) are very mild, as they are both satisfied
whenever
\begin{eqnarray}h_n n^{-\zeta}+h_n^{-1}n^{\eta}\rightarrow0\label{bandwidth3},\;
\mbox{for some}\;-1/2<\eta\leq\zeta<0.\end{eqnarray}\vskip6pt

Hence we shall be able to choose between the following methods for
approximating quantiles of the distribution of $T_{\cal
S}(P_n,\Gamma,\nu)$ and constructing rejection regions for our
test statistic:

\begin{itemize}\item We can simulate the
Brownian bridge and compute the quantiles of the distribution of
its supremum over the data dependent class $\hat{\cal S}_{b,h_n}$
for some choice of $h_n$.\item We can use a subsampling
approximation of the quantiles of the distribution of $T_{\cal
S}(P_n,\Gamma,\nu)$. Indeed,
$\sup_{A\in{\cal S}_b}\mathbb{G}(A)$ has continuous distribution
function on $[0,+\infty)$, hence the subsampling approximation of
quantiles is valid.
\end{itemize}

Before moving on to specific asymptotic results, we close this
heuristic description with a discussion of the cases where the
class of saturated sets ${\cal S}_b$ is the trivial class
$\{\varnothing,{\cal Y}\}$. In such cases, the test statistic
converges to zero if one chooses the scaling factor $\sqrt{n}$. A
refinement of the test will therefore involve a faster rate of
convergence, determined through the construction of a local
empirical process taylored to the shape of $\nu\Gamma$ close to
$\varnothing$ and to ${\cal Y}$.

\subsubsection{Specific asymptotic results} We now turn to
specific conditions on the structure $(\Gamma,\nu)$ and the law
$P$ of the observables such that results (\ref{asymptotic
distribution}) which allows the subsampling approach, and
(\ref{asymptotic approximation}) which then also allows the
simulation approach, hold. \begin{itemize}\item[(a)] Case where
${\cal Y}$ is finite and ${\cal S}$ is the class of all subsets
${\cal S}=2^{\cal Y}$.\vskip4pt In that case, we show in
Theorem~3a below that both approaches to inference are valid.

\vskip8pt\noindent{\bf Theorem~3a:} If ${\cal Y}$ is finite and
${\cal S}=2^{\cal Y}$, (\ref{asymptotic distribution}) and
(\ref{asymptotic approximation}) hold. \item[(b)] Case where
${\cal Y}=\mathbb{R}^{d_y}$, $P$ is absolutely continuous with
respect to Lebesgue measure and ${\cal
S}=\{(y_1,z_1)\times\ldots\times(y_{d_{y}},z_{d_{y}}):
\;y_1,\ldots,y_{d_{y}},z_1,\ldots,
z_{d_y}\in\overline{\mathbb{R}}\}$ or any subclass, such as the
class ${\cal C}$ defined above\footnote{Note that since $P$ is
absolutely continuous, considering only open intervals is without
loss of generality.}.\vskip4pt

As indicated above, the asymptotic results are derived under
assumptions such that the function $\delta$ ``takes off'' frankly
from zero. To make this precise, we introduce the following
``frank separation'' assumption. Recall that if $d$ is the
Euclidean metric on ${\cal Y}$, the Haussdorf metric
$d_H$ between two sets $A_1$ and $A_2$
is defined by
\begin{eqnarray*}d_H(A_1,A_2)=\max\left(
\sup_{y\in A_1}\inf_{z\in A_2}d(y,z),\sup_{z\in A_2}\inf_{y\in
A_1}d(y,z)\right).\end{eqnarray*} We need to ensure that on sets
that are sufficiently distant from sets in ${\cal S}_b$ (where the
inequality is binding), then $\delta$ is sufficiently negative so
that it dominates local oscillations of the empirical process. To
formalize this, we define the subclass of ${\cal S}$ of sets such
that the inequality is nearly binding.

\vskip8pt\noindent{\bf Definition 3.3:} We denote the subclass of
sets from ${\cal S}$ where $P\geq\nu\Gamma-h$ by ${\cal S}_{b,h}$,
i.e. \begin{eqnarray*}{\cal S}_{b,h}:=\left\{A\in{\cal S}:\;
P(A)\geq\nu(\Gamma(A))-h\right\}.\end{eqnarray*}

We can now state \vskip8pt\noindent{\bf Assumption FS (Frank
Separation):} There exists $K>0$ and $0<\eta<1$ such that for all
$A\in{\cal S}_{b,h}$, for $h>0$ sufficiently small, there exists
an $A_b\in{\cal S}_b$ such that $A_b\subseteq A$ and
$d_H(A,A_b)\leq K h^{\eta}$.\vskip6pt

\noindent{\bf Remark 1:} Assumption is very mild, in the sense
that it fails only in pathological cases, such as the case where
${\cal Y}=\mathbb{R}$, ${\cal S}={\cal C}$, and $y\mapsto
P((-\infty,y])-\nu(\Gamma((-\infty,y]))$ is $C^{\infty}$ with all
derivatives equal to zero at some $y=y_0$ such that
$(-\infty,y_0]\in{\cal C}$. \vskip4pt

Then, we have:

\vskip8pt\noindent{\bf Theorem~3b:} Suppose assumptions~FS and
(\ref{bandwidth3}) hold and that $P$ is absolutely continuous with
respect to Lebesgue measure. Then (\ref{asymptotic distribution})
and (\ref{asymptotic approximation}) hold.\vskip6pt

The proof is based on the following lemma,

\vskip8pt\noindent{\bf Lemma~3a:} Under the conditions of
Theorem~3b, we have
\begin{eqnarray*}\sup_{A\in{\cal S}_{b,h_n}}\mathbb{G}_n(A)
\rightsquigarrow\sup_{A\in{\cal S}_b}\mathbb{G}(A),\end{eqnarray*}

which involves bounds on oscillations of the empirical
process.\vskip6pt
\end{itemize}

\subsection{Power of the test}
As mentioned before, to ensure consistency of our specification
test statistic, we need to derive conditions on the structure
$(\Gamma,\nu)$ and the law $P$ of observables such that all
violations of the inequality $P\leq\nu\Gamma$ will be detected
asymptotically with a test based on the statistic $T_{\cal
S}(P_n,\Gamma,\nu)$.\vskip6pt Before giving specific results, we
shall try to convey the extent of the difficulties involved, in
comparison with the case of the classical Kolmogorov-Smirnov test
which was developed in our prelude.\vskip6pt When testing the
equality of two probability measures, as in the Kolmogorov-Smirnov
test, we need a class of sets that will determine the value of the
law $P$, since it will ensure that if the equality holds on this
class of sets, it holds everywhere. To be more precise, we need a
convergence determining class (see section 2.6 page 18 of
\cite{Vaart:98}) since our test is asymptotic.\vskip6pt When
testing the inequality $P\leq\nu\Gamma$, the situation is
complicated in two ways. First, $\nu\Gamma$ is a set function, but
it is generally not additive unless $\Gamma$ is bijective, and a
convergence determining class is much harder come by. Second,
determining the value of $\nu\Gamma$ may not be sufficient, since
it may not guarantee that the direction of the inequality
$P\leq\nu\Gamma$ will be maintained from the reduced convergence
determining class to all measurable sets. We discuss these two
points in the following subsections.

\subsubsection{Convergence determining classes for $\nu\Gamma$:}
The set function $A\mapsto\nu(\Gamma(A))$ is a Choquet capacity
functional (for definitions and properties, see Appendix~A2), and
the following lemma (lemma 1.14 of \cite{SW:86}) provides a
convergence determining class in great generality. Recall that a
closed ball $B(y,\eta)$ with center $y$ and radius $\eta$ is the
sets of points in ${\cal Y}$ whose distance to $y$ is lower or
equal to $\eta$. Define ${\cal S}_{{\mathrm{SW}}}$ as the class of
compact subsets of ${\cal Y}$ with the following two properties:
\begin{itemize}\item[(C1)] Elements of ${\cal
S}_{{\mathrm{SW}}}$ are finite unions of closed balls with
positive radii, \item[(C2)] Elements of ${\cal S}_{{\mathrm{SW}}}$
are continuity sets for the Choquet capacity functional
\begin{eqnarray*}A\rightarrow\nu(\Gamma(A)),
\end{eqnarray*} in other words, if $A\in{\cal S}_{{\mathrm{SW}}}$,
then
$\nu(\Gamma(\mbox{cl}(A)))=\nu(\Gamma(\mbox{int}(A)))$.\end{itemize}
Then we have:

\vskip8pt\noindent{\bf Lemma SW:} The class ${\cal
S}_{{\mathrm{SW}}}$ is convergence determining.\vskip6pt

The class ${\cal S}_{{\mathrm{SW}}}$ is not a
Vapnik-$\breve{\mbox{C}}$ervonenkis class of sets since for any
finite collection of points, there is a collection of finite union
of balls that shatters it (see appendix~A1). However, there is a
natural restriction of this class which is. In the case where
${\cal Y}=\mathbb{R}^{d_y}$, ${\cal S}_{{\mathrm{SW}}}$ can be
redefined with rectangles instead of balls. Take an integer $K$.
Define the class of finite unions of {\em at most} $K$ rectangles:
\begin{eqnarray*}{\cal S}_K&=&\{\,\bigcup_{k\leq
K}(y_k,z_k):\;(y_k,z_k)\in \mathbb{R}^{2d_y}\}.
\end{eqnarray*} Then we have

\vskip8pt\noindent{\bf Lemma~3b:} \hskip4pt ${\cal S}_K$ is a
Vapnik-$\breve{\mbox{C}}$ervonenkis class of sets.\vskip6pt

Hence this class is amenable to asymptotic treatment.

\subsubsection{Core determining classes for $\nu\Gamma$}
The requirement, that we call ``Core determining'', on the class
${\cal S}$ that $P(A)\leq\nu(\Gamma(A))$ for all $A\in{\cal S}$
imply $P(A)\leq\nu(\Gamma(A))$ for all measurable $A$ is
apparently more stringent than the requirement that the values of
the set function $\nu(\Gamma(.))$ on all measurable sets be
determined by its values on ${\cal S}$.

\vskip8pt\noindent{\bf Definition 3.4:} A class ${\cal S}$ of
subsets of ${\cal Y}$ is core determining for $(\Gamma,\nu)$ if
\[ \sup_{\cal S}\,(P-\nu\Gamma)=0\;\Longrightarrow\;\sup_{{\cal
B}_{\cal Y}}\,(P-\nu\Gamma)=0\]\vskip6pt

We have noted already the obvious fact:

\vskip8pt\noindent{\bf Fact 1:} ${\cal S}=2^{\cal Y}$ is core
determining for observables on a finite set ${\cal Y}$.\vskip6pt

A close inspection of the proof of Theorem~2 shows the following
fact:

\vskip8pt\noindent{\bf Fact 2:} The class ${\cal F}_{\cal Y}$ of
closed subsets of ${\cal Y}$ is core determining.\vskip6pt

We now show that we can actually say much more by linking the core
determining property with the convergence determining property,
and showing that the class $\tilde{{\cal S}}_{{\mathrm{SW}}}$ of
finite unions of open balls with positive raddii (or alternatively
the class finite unions of open rectangles) is core
determining.\vskip4pt

First, we need to consider the following assumptions on the
structure:

\vskip8pt\noindent{\bf Assumption (CD1):} ${\cal Y}$ is a compact
subset of $\mathbb{R}^{d_y}$, and ${\cal U}$ is a compact subset
of $\mathbb{R}^{d_u}$.\vskip4pt

\vskip8pt\noindent{\bf Assumption (CD2):} $P$ and $\nu$ are
absolutely continuous with respect to Lebesgue measure.\vskip4pt

\vskip8pt\noindent{\bf Assumption (CD3):} There exists
$\gamma_0\in$ Sel$(\Gamma)$ such that $P(A)\rightarrow0$ implies
$\nu(\gamma_0(A))\rightarrow0$.\vskip4pt

Note that assumption~(CD3) is satisfied if either of the following
hold: \begin{itemize}\item There exists $\gamma_0\in$
Sel$(\Gamma)$ injective, such that $\nu\gamma_0$ (now a
probability measure) is absolutely continuous with respect to $P$.
\item There exists $\gamma_0\in$ Sel$(\Gamma)$ and $\alpha>0$ such
that $\nu(\gamma_0(A))\leq \alpha P(A)$ for all $A$
measurable.\end{itemize}

\vskip8pt\noindent{\bf Assumption (CD4):} $\Gamma$ is
convex-valued, i.e. $\Gamma(y)$ is a convex set for all $y\in{\cal
Y}$.\vskip4pt

This assumption rules out some interesting cases, for instance
when the graph of $\Gamma$ (defined in (\ref{graph})) is the union
of the graphs of two functions. However, our conditions are not
minimal, and such cases could be treated under a different set of
conditions.\vskip6pt

We define the upper and lower envelopes of the Graph of $\Gamma$
by

\vskip8pt\noindent{\bf Definition 3.5:} The upper (resp. lower)
envelope of Graph $\Gamma$ is the function $y\mapsto
u(y)=\sup\,\{\Gamma(y)\}$ (resp. $y\mapsto l(y)=\inf\,
\{\Gamma(y)\}$).\vskip4pt

\vskip8pt\noindent{\bf Assumption (CD5):} The upper and lower
envelopes $u$ and $l$ of the graph of $\Gamma$ are Lipschitz, i.e.
there exists $\kappa\geq0$ such that for all $y_1,y_2\in{\cal Y}$,
\[\max\left(\vert u(y_1)-u(y_2)\vert,\vert
l(y_1)-l(y_2)\vert\right)\leq \kappa\vert y_1-y_2\vert.\]\vskip4pt

To state our last assumption, we need an extra definition:

\vskip8pt\noindent{\bf Definition~3.6:} A forking point of
$\Gamma$ is a $y_0$ such that for any $\epsilon>0$, there exists
$y_1$ and $y_2$ in the open ball B$(y_0,\epsilon)$ such that
$\Gamma(y_1)$ is a singleton, and $\Gamma(y_2)$ is not.\vskip4pt

\vskip8pt\noindent{\bf Assumption (CD6):} $\Gamma$ has at most a
finite number of forking points.\vskip4pt

Note that this is a technical assumption that is violated only in
pathological cases, and that is akin to the Frank Separation
Assumption (FS).

We can now state the result:

\vskip8pt\noindent{\bf Theorem~3c:} Under assumption~(CD1)-(CD6),
the class $\tilde{{\cal S}}_{{\mathrm{SW}}}$ of finite unions of
open balls with positive radii (or alternatively the class finite
unions of open rectangles) is core determining.\vskip6pt

This result is fundamental in that it reduces the problem of
checking consistency of the test based on the statistic $T_{\cal
S}(P_n,\Gamma,\nu)$ to the problem of checking whether
$P(A)\leq\nu(\Gamma(A))$ for $A$ a finite union of balls (or
rectangles) in $\mathbb{R}^{d_y}$ whenever $P\leq\nu\Gamma$ on
${\cal S}$.

\begin{figure}[htbp]
\begin{center}
\includegraphics[width=8cm]{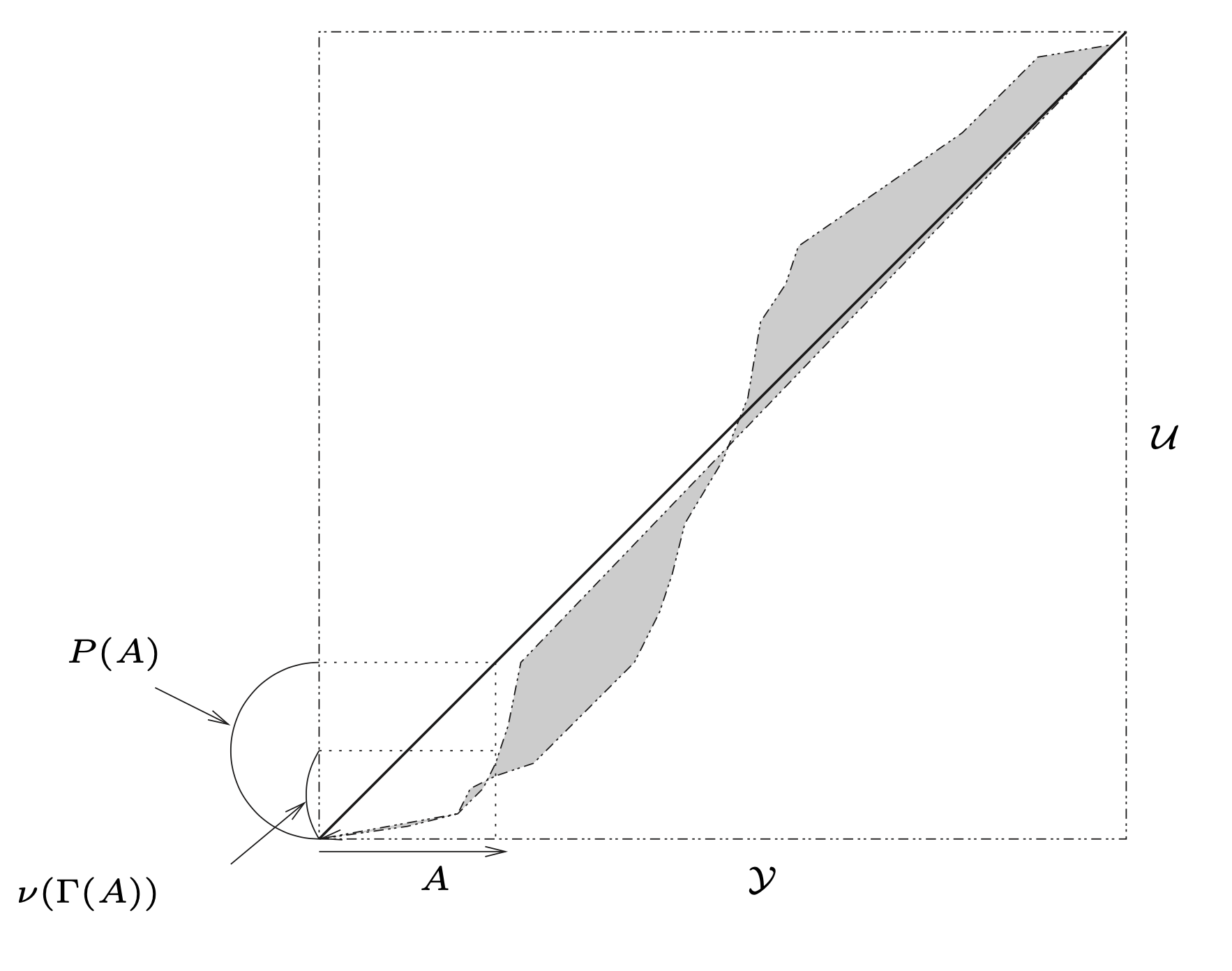}
\caption{Violation of null that can be detected by the class of
cells ${\cal C}$. Notice in particular that the inequality $P\leq
\nu\Gamma$ is violated on the set $A$ ($P$ and $\nu$ are
uniform).} \label{detected}
\end{center}
\end{figure}

We shall now apply this reasoning to give some conditions on the
structure $(\Gamma,\nu)$ under which the test based on statistic
$T_{\cal S}(P_n,\Gamma,\nu)$ is consistent with ${\cal S} = {\cal
C} = \{(-\infty,y], (y,\infty): \;y\in\mathbb{R}\}$, such as in
figure~\ref{detected}, and conditions under which the class ${\cal
C}$ may not be core determining, but the class ${\cal S} = {\cal
R} = \{(y,z): \;y,z\in\overline{\mathbb{R}}\}$ is. We thereby
defining classes of alternatives that our tests based on $T_{\cal
C}(P_n,\Gamma,\nu)$ and $T_{\cal R}(P_n,\Gamma,\nu)$ have power
against in case ${\cal Y}=\mathbb{R}$ and $P$ is absolutely
continuous with respect to Lebesgue measure.\vskip6pt

\vskip6pt

\vskip8pt\noindent{\bf Theorem~3d:} If assumption~(CD1) and (CD2)
are satisfied, and the graph of $\Gamma$ has increasing upper and
lower envelopes, then ${\cal C}$ is core determining, and hence
the specification test based on the statistic $T_{\cal
C}(P_n,\Gamma,\nu)$ is consistent.\vskip6pt

\begin{figure}[htbp]
\begin{center}
\includegraphics[width=8cm]{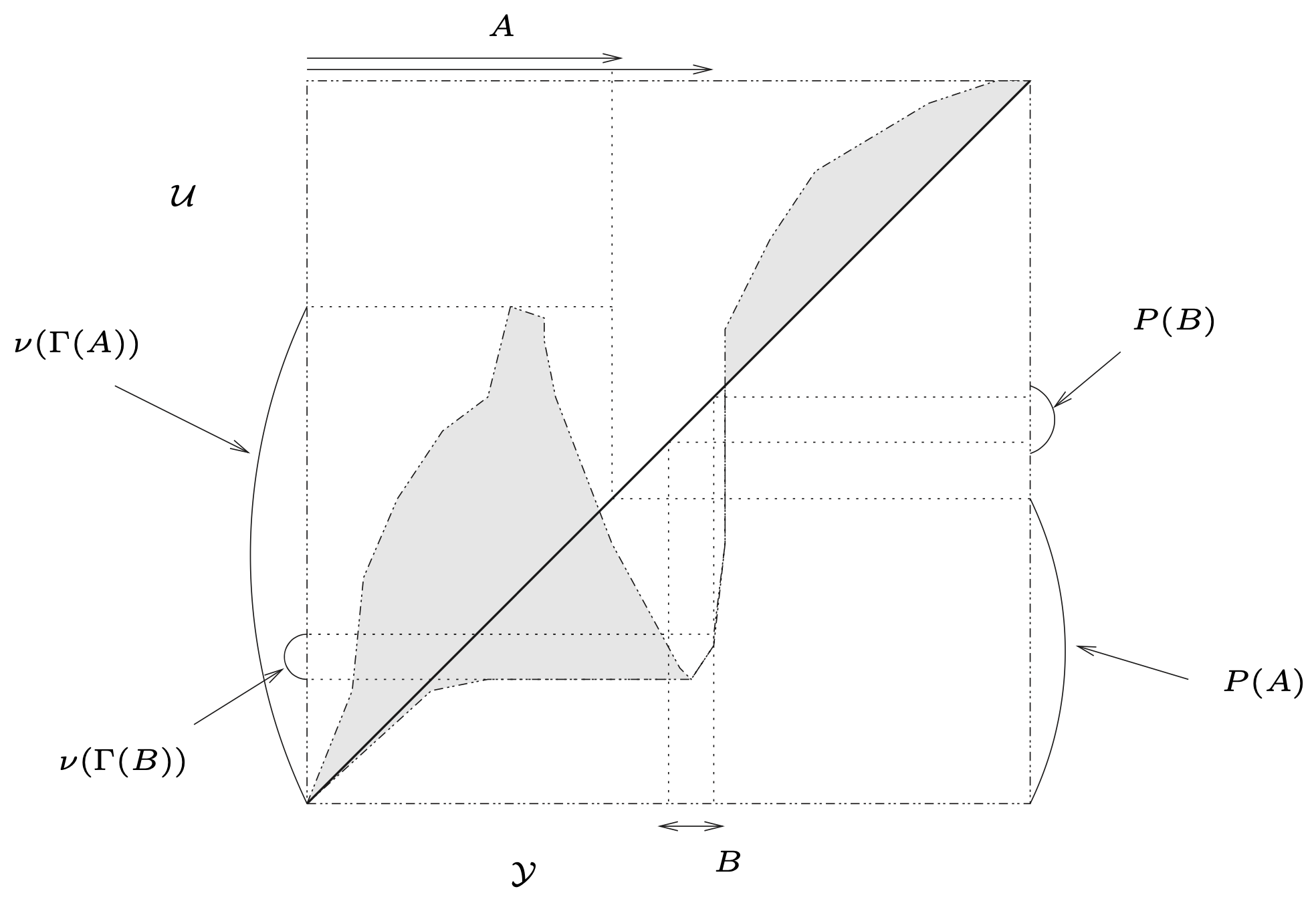}
\caption{Violation of null that cannot be detected by the class of
cells ${\cal C}$, but can be detected by the class of all
intervals. Notice in particular that the inequality $P\leq
\nu\Gamma$ is violated on $A$ but not on $B$ ($P$ and $\nu$ are
uniform).} \label{undetected}
\end{center}
\end{figure}

In figure~\ref{undetected}, we show a case where the null
hypothesis does not hold, but a test based on $T_{\cal
C}(P_n,\Gamma,\nu)$ fails to detect it because of the lack of
monotonicity of the upper envelope. In that case, we need the
larger class of sets ${\cal R}$ to detect the departure from the
null.

\section{Applications of the inference framework}
The test of specification that we have developed can be applied to
the construction of confidence regions in case the structure
depends on unknown parameters. Let
$\theta\in\Theta\subseteq\mathbb{R}^{d_\theta}$ be a vector of
structural parameters, and let the model be given by
$(\Gamma_{\theta},\nu_{\theta})$.

\vskip8pt\noindent{\bf Definition 4.1:} The identified set
$\Theta_I$ is defined as the set of all $\theta\in\Theta$ such
that the null hypothesis H$_0(\theta)$ of compatibility of
$(\Gamma_{\theta},\nu_{\theta})$ with $P$ (as defined in
Theorems~1 and~1') holds true. \vskip4pt

This section is an outline of the application of our testing
procedure to the construction of confidence regions for elements
of the identified set and for the identified set itself.\vskip4pt

\subsection{Coverage of parameters in the identified set}
To form a confidence region that covers (with at least some
pre-determined probability) each parameter value that makes the
structure compatible with the distribution of observables, we
propose to invert our test statistic to form a confidence region
for elements of $\Theta_I$. In other words, for a given
$\alpha\in(0,1)$, we seek a region CR$_n$ such that, for all
$\theta\in\Theta_I$, $\liminf_{n}\mathbb{P}(\theta\in\mbox{CR}_n)
\geq \alpha.$ The confidence region obtained from inverting the
test has the form CR$_n = \{\theta\in\Theta: \sqrt{n} T_{\cal S}
(P_n,\Gamma_{\theta},\nu_{\theta}) \leq
\hat{Q}_{\alpha}(\theta)\}$ where ${\cal S}$ is a class of sets
which is Core determining for all $\theta\in\Theta$ and
$\hat{Q}_{\alpha}(\theta)$ is an approximation of the $\alpha$
quantile of the distribution of $T_{\cal S}
(P_n,\Gamma_{\theta},\nu_{\theta})$. A valid approximation can be
obtained using either one of the two methods proposed at the end
of section~3.1.1.

\subsection{Coverage of the identified set} To form a region that
covers the whole identified set with pre-determined probability,
we need a region $\mbox{CR}^*_n$ such that
$\liminf_{n}\mathbb{P}(\Theta_I\subseteq\mbox{CR}^*_n) \geq
\alpha.$ The latter can be obtained using the method proposed by
\cite{CHT:2002} applied to the criterion function
$(\sup_{A\in\cal S} (P(A)-\nu_{\theta}(\Gamma_{\theta}(A))))^2$
with sample criterion $T^2_{\cal S}
(P_n,\Gamma_{\theta},\nu_{\theta})$ (under the condition that C1,
C2, C4 and C5 of \cite{CHT:2002} hold). A main contribution
of this paper, therefore, is to provide the first natural and
general choice of criterion function, and thereby pave the way for
a comparison of criteria and a discussion of optimality.

\subsection{Illustration}
We now spell out our procedures on a very simple example:
example~5 of section~1. The structure is described by the
multi-valued mapping: $\Gamma(1)=[0,\lambda]$ and
$\Gamma(0)=[0,1]$. In this case, since $y$ is Bernoulli, we can
write $P=(1-p,p)'$ with $p$ the probability of a 1. For the
distribution of $u$, we consider a parametric exponential family
on $[0,1]$. Hence $\nu_{\phi}$ has distribution function
$u^{\phi}$, with $\phi>0$. Our parameter vector is therefore
$\theta = (\lambda,\phi)'$.\vskip4pt

The null hypothesis in this case is immediately seen to be
equivalent to $p\leq\lambda^{\phi}$ for a given value of the
parameter vector. Indeed, the easiest formulation to use is
probably formulation (v) which requires that $p= P(\{1\}) \leq
\nu(\Gamma(1)) = \nu[0,\lambda] = \lambda^{\phi}$. Hence
$T_{2^{\{0,1\}}}(P_n,\Gamma_{\theta},\nu_{\theta}) =
p_n-\lambda^{\phi}$. Now, if $p = \lambda^{\phi}$, then ${\cal
S}_b = \{ \varnothing, \{0\}, \{1\}, \{0,1\}\}$ and then
$\sqrt{n}(p_n-\lambda^{\phi})$ converges weakly to a normal random
variable with mean zero and variance $p(1-p)$, whereas if $p <
\lambda^{\phi}$, then ${\cal S}_b = \{ \varnothing, \{0,1\}\}$ and
$\sqrt{n}(p_n-\lambda^{\phi})$ converges to zero. In either case,
for a given choice of sequence $h_n$, $\hat{\cal S}_{b,h_n}$ is
equal to $\{ \varnothing, \{0\}, \{1\}, \{0,1\}\}$ if $p_n \geq
\lambda^{\phi} - h_n$ and $\{ \varnothing, \{0,1\} \}$
otherwise.\vskip4pt

The $\alpha$ quantile of $\sqrt{n}
T_{2^{\{0,1\}}}(P_n,\Gamma_{\theta},\nu_{\theta}) = \sqrt{n}( p_n
- \lambda^{\phi})$ can be approximated with 0 if $p_n <
\lambda^{\phi} - h_n$, and with the $\alpha$ quantile of the
normal with mean zero and variance $p_n(1-p_n)$ if $p_n \geq
\lambda^{\phi} - h_n$. Alternatively, $Q_{\alpha}(\theta)$ can be
approximated using subsampling (though it would be a serious case
of overkill). The procedure would then be the following: Consider
all (or a large number $B_n$ of) the samples of size $b_n$ from
the sample of size $n$ with $1/b_n + b_n/n \rightarrow 0$ and
approximate $Q_{\alpha}(\theta)$ with
\[ \hat{Q}_{\alpha}(\theta) = \inf\{ x  :\, \frac{1}{B_n}\sum_{i=1}^{B_n} \{\sqrt{b}
T_{\cal S}(P_b^i,\Gamma_{\theta},\nu_{\theta}) \leq x\} \geq
\alpha \} \] where $P_b^i$ is the empirical distribution of the
$i$-th subsample. A confidence region is then CR$_n =
\{\theta\in[0,1]\times(0,+\infty): \sqrt{n} T_{\cal S}
(P_n,\Gamma_{\theta},\nu_{\theta}) \leq
\hat{Q}_{\alpha}(\theta)\}$.

\subsection{Semi-nonparametric extensions}

Since structures are often given without a specification of the
distribution of the unobservable variables, it is customary to
assume only moment conditions, such as a given mean (taken to be
equal to zero without loss of generality) and finite variance.
This includes as special cases structures defined by moment
inequality conditions.\vskip6pt

In such cases, a similar approach can be taken where the null is
defined as the existence of a joint law supported on the set
$\{u\in\Gamma_{\theta}(y)\}$ with marginal $P$ on ${\cal Y}$ and
marginal on ${\cal U}$ satisfying some moment conditions. Calling
${\cal V}$ the set of laws that satisfy the said conditions, the
dual formulation delivers a feasible version of the statistic
\[\inf_{\nu\in{\cal V}}\sup_{A\in{\cal
S}}\,\left[P(A)-\nu(\Gamma_{\theta}(A))\right].\] This involves a
number of difficulties, which are the subject of a companion paper
\cite{GH:2006}. We only give here, as an illustration, the
application of the method on a classic special case of
example~3\vskip6pt

Suppose one observes income brackets with centers in ${\cal
Y}=\{y_1,\ldots,y_k\}$ with $y_1<\ldots<y_k$ and width $\delta$.
True income is unobservable, and one is interested in the mean of
true income. The model correspondence is given by
$\Gamma(y)=(y-\delta/2,y+\delta/2)$. Let $p(y_i)$ (resp.
$p_n(y_i)$) denote the true (resp. empirical) probability of
$\{Y=y_i\}$.\vskip6pt

Consider formulation (v'): $\nu\leq P\Gamma^{-1}$ of the null
hypothesis. Denoting $\Gamma^{u}(B)=\{y:\;\Gamma(y)\subseteq B \}$
for any $B\in{\cal B}_{\cal U}$, and writing $\phi^*=P\Gamma^{-1}$
and $\phi_*=P\Gamma^{u}$, we have (using Definition~A2.6
Lemma~A2.2 in appendix~A2) that under the null, the expectation of
any measurable function $f$ of the unobservable variables
satisfies
\[\int_{{\mathrm{Ch}}}f\,d\phi_*\leq
\mathbb{E}f\leq\int_{{\mathrm{Ch}}}f\,d\phi^*.\] Denoting
$\phi_n^*=P_n\Gamma^{-1}$ and $\phi_{n*}=P_n\Gamma^{u}$ the
empirical versions of $\phi^*$ and $\phi_*$, the set
$[\int_{{\mathrm{Ch}}}f\,d\phi_{n*},\int_{{\mathrm{Ch}}}f\,d\phi_n^*]$
estimates the identified set $[\int_{{\mathrm{Ch}}}f\,d\phi_*
\int_{{\mathrm{Ch}}}f\,d\phi^*].$ In the case considered here,
where $f$ is the identity, this identified set equals
\[\left[\sum_{i=1}^k (y_i-\delta/2)\,p(y_i),\sum_{i=1}^k
(y_i+\delta/2)\,p(y_i)\right],\] which is equal to
\[\left[\sum_{i=1}^k
(y_i-\delta/2)\,(p_n(y_i)-g_{n,i}/\sqrt{n}),\sum_{i=1}^k
(y_i+\delta/2)\,(p_n(y_i)-g_{n,i}/\sqrt{n})\right]\] from which
asymptotically valid confidence regions can be constructed, since
$g_n = (g_{n,1}$, $\ldots$, $g_{n,k})'$, with $g_{n,i} =
\sqrt{n}(p_n(y_i)-p(y_i))$ is asymptotically a Gaussian vector.

\section*{Conclusion}
We have provided a coherent definition of correct specification of
structures with no identifying assumptions. This definition is the
result of the equivalence of several natural generalizations of
the hypothesis of correct specification in the identified case.
These theoretical formulations of correct specification have
natural empirical counterparts, several of which are also shown to
be equivalent, and a test of specification is based on the latter.
When the structure is parameterized, this test can be inverted to
yield confidence regions for the set of structural parameters for
which the null hypothesis of correct specification is
satisfied.\vskip6pt

This work has the following natural extensions: First, the whole
approach is articulated around the existence of a joint measure
with given marginals, hence it is essentially parametric in
nature, but can be naturally extended to a problem of existence of
a joint probability measure with one marginal given (the
distribution of observables) and moment conditions on the other
marginal (the distribution of unbobservable variables). This
natural extension of our work will nest structures defined by
moment inequalities, and therefore deliver a way to construct
confidence regions in such cases. Second, the statistic we have
used to examine correct specification can be derived from the
Kolmogorov-Smirnov distance between the empirical distribution and
the set of data generating processes implied by the structure.
Other distances and pseudo-distances will generate different
specification statistics, and relative entropy may be a
particularly good candidate, in that it produces optimal inference
in the special case of identified structures.

{\small

\subsection*{Appendix A: Additional concepts and results}

\subsubsection*{A1: Convergence of the empirical process}
We give here definitions and results that we use in our asymptotic
analysis. The definition of a Vapnik-$\breve{\mbox{C}}$ervonenkis
class of sets is given in section 2.6.1 page 134 of
\cite{VW:96} and reproduced here for the convenience of the
reader.

\vskip8pt\noindent{\bf Definition~A1.1:} Let ${\cal S}$ be a
collection of subsets of a set ${\cal X}$. An arbitrary set of $n$
points $\{x_1,\ldots,x_n\}$ posesses $2^n$ subsets. Say that
${\cal C}$ picks out a certain subset from $\{x_1,\ldots,x_n\}$ if
this can be formed as the set $C\cap\{x_1,\ldots,x_n\}$ for a $C$
in ${\cal S}$. The collection ${\cal S}$ is said to shatter
$\{x_1,\ldots,x_n\}$ if each of its $2^n$ subsets can be picked
out in this manner. The Vapnik-$\breve{\mbox{C}}$ervonenkis index
of the class ${\cal S}$ is the smallest $n$ for which no set of
cardinality $n$ is shattered by ${\cal S}$. A
Vapnik-$\breve{\mbox{C}}$ervonenkis class of sets is a class with
finite Vapnik-$\breve{\mbox{C}}$ervonenkis index.

\vskip8pt\noindent{\bf Fact~A1:} The class of cells ${\cal C}$ is
a Vapnik-$\breve{\mbox{C}}$ervonenkis class of sets (see Example
2.6.1 page 135 of \cite{VW:96}).

\vskip8pt\noindent{\bf Definition~A1.2:} The $P$-Brownian bridge
is the tight centered Gaussian stochastic process with
variance-covariance defined by
$\mathbb{E}\mathbb{G}(A_1)\mathbb{G}(A_2)=P(A_1\cap A_2)
-P(A_1)P(A_2)$.

\vskip8pt\noindent{\bf Theorem~A1.1:} If ${\cal S}$ is a
Vapnik-$\breve{\mbox{C}}$ervonenkis class of sets, the empirical
process converges weakly to the $P$-Browninan bridge $\mathbb{G}$,
and the convergence is uniform over the class ${\cal S}$ (i.e. the
convergence is in $l^{\infty}({\cal F})$, where ${\cal F}$ is the
class of indicator functions of sets in ${\cal S}$).

\vskip6pt\noindent{\bf Proof of Theorem~A1.1:} We assume that
${\cal S}$ is a Vapnik-$\breve{\mbox{C}}$ervonenkis class of sets.
Call ${\cal F}$ the class of indicator functions of sets in ${\cal
S}$, and call $V({\cal F})$ the
Vapnik-$\breve{\mbox{C}}$ervonenkis index of the corresponding
class of sets. By Theorem 2.6.4 page 136, there exists a constant
$C$ such that for all probability measure $Q$ and all
$0<\varepsilon<1$, the covering number (see definition 2.2.3 page
98 of \cite{VW:96}) of ${\cal F}$ in $\mathbb{L}_2(Q)$
metric, N$(\varepsilon,{\cal F},\mathbb{L}_2(Q))$ satisfy
\begin{eqnarray*}\mbox{N}(\varepsilon,{\cal F},\mathbb{L}_2(Q))
\leq C(V({\cal F}))(4e)^{V({\cal F})}(1/\varepsilon)^{2(V({\cal
F})-1)}.\end{eqnarray*} Hence, we have
\begin{eqnarray*}\int_0^{\infty}\sup_{Q}\sqrt{\ln\mbox{N}
(\varepsilon,{\cal
F},\mathbb{L}_2(Q))}\;d\varepsilon<\infty.\end{eqnarray*} Since
${\cal F}$ is a class of indicator functions, the above suffices
to satisfy conditions of Theorem 2.5.2 page 127 of
\cite{VW:96}, and ${\cal F}$ is $P$-Donsker, which by
definition means that $\mathbb{G}_n$ converges in
$l^{\infty}({\cal F})$.\vskip6pt

By the continuous mapping theorem, we immediately have the
following corollary:

\vskip6pt\noindent{\bf Corollary~A1.1:} If ${\cal S}$ is a
Vapnik-$\breve{\mbox{C}}$ervonenkis class of sets, then
$\sup_{\cal S}\mathbb{G}_n$ converges weakly to $\sup_{\cal
S}\mathbb{G}$.

\subsubsection*{A2: Choquet capacity functionals}
We collect here all the definitions, equivalent representations
and properties of Choquet capacity functionals (a.k.a.
distributions of random sets or infinitely alternating capacities)
that are useful for this paper. All the results presented here can
be traced back to \cite{Choquet:53}.\vskip4pt

Take ${\cal X}$ a Polish space (complete metrizable and seperable
topological space) endowed with its Borel $\sigma$-algebra ${\cal
B}$. For a sequence of numbers, $a_n\uparrow a$ (resp.
$a_n\downarrow a$) denotes convergence in inceasing (resp.
decreasing) values, whereas for a sequence of sets, the notation
$A_n\uparrow A$ (resp. $A_n\downarrow A$) denotes $A_n\subseteq
A_{n+1}$ for all $n$ and $A=\bigcup_{n}A_n$ (resp.
$A_{n+1}\subseteq A_{n}$ for all $n$ and $A=\bigcap_{n}A_n$).
Finally, denote ${\cal F}$ (resp. ${\cal G}$) the set of closed
(resp. open) subsets of ${\cal X}$, and for $A\in{\cal B}$, ${\cal
F}_A = \{F\in{\cal F}:\;F\cap A\neq\varnothing\}$.

\vskip8pt\noindent{\bf Definition A2.1:} A capacity is a set
function $\varphi:\,{\cal B}\rightarrow \mathbb{R}$ satisfying
\begin{itemize}\item[(i)] $\varphi(\varnothing)=0$ and $\varphi({\cal X})=1$,
\item[(ii)] For any two Borel sets $A\subseteq B$, we have
$\varphi(A) \leq \varphi(B)$, \item[(iii)] For all sequences of
Borel sets $A_n \uparrow A$, we have $\varphi(A_n) \uparrow
\varphi(A)$, \item[(iv)] For all sequences of closed sets $F_n
\downarrow F$, we have $\varphi(F_n) \downarrow
\varphi(F)$.\end{itemize}

\vskip8pt\noindent{\bf Definition~A2.2} A capacity $\varphi$ is
called infinitely alternating if for any $n$ and any sequence
$A_1,\ldots,A_n$ of Borel sets,
\begin{eqnarray*}\varphi\left(\bigcap_{i=1}^{n}A_i\right)
\leq\sum_{\varnothing\neq I\subseteq\{1,2,\ldots,n\}}
(-1)^{|I|+1}\varphi\left(\bigcup_I A_i\right)\end{eqnarray*}

We call Choquet capacity functional an infinitely alternating
capacity. Probability measures are special cases of Choquet
capacity functionals, for which the alternating inequality of
definition~A2.2 holds as an equality (known as Poincar\'e's
equality).\vskip4pt

We now show that infinite alternation is a characteristic property
of distributions of random sets (for a proof, see for instance
section 2.1 of \cite{Matheron:75}).

\vskip8pt\noindent{\bf Theorem~A2.1:} $\varphi$ is a Choquet
capacity functional (i.e. an infinitely alternating capacity) if
and only if there exists a probability measure ${\cal P}$ on
${\cal F}$ such that, for all $A\in{\cal B}$, $\varphi(A)=$ ${\cal
P}({\cal F}_A)$, and such a ${\cal P}$ is unique.\vskip4pt

$\varphi$ is therefore called the distribution of the random set
associated with the probability measure ${\cal P}$, which allows
the following definition of convergence determining classes for a
Choquet capacity functional:

\vskip8pt\noindent{\bf Definition~A2.3:} A class ${\cal C}$ of
Borel subsets of ${\cal X}$ is called convergence determining for
a Choquet capacity functional $\varphi$ if and only if the class
$\{{\cal F}_A\,;\;A\in{\cal C}\}$ is convergence determining for
the probability measure ${\cal P}$ associated to $\varphi$ as in
Theorem~A2.1.\vskip4pt

We now look at the relation with measurable correspondences,
defined as correspondences that satisfy Assumption~1 in the main
text. Let $(\Omega,{\cal B},\mathbb{P})$ be a probability space.

\vskip8pt\noindent {\bf Definition~A2.4:} A non-empty and closed
valued correspondence $\Gamma:$ $\Omega\rightrightarrows{\cal X}$
is called a measurable correspondence if for each open set ${\cal
O}\subseteq {\cal X}$, $\;\Gamma^{-1}({\cal
O})=\{\omega\in\Omega\;|\; \Gamma(\omega)\cap{\cal
O}\neq\varnothing\}$ belongs to ${\cal B}$.\vskip6pt

If we define $\varphi$ by
$\varphi(A)=\mathbb{P}\{\omega\in\Omega\;| \;\Gamma(\omega)\cap
A\neq\varnothing\}$, for all $A\in{\cal B}$, then $\varphi$ is a
Choquet capacity functional (from section 26.8 page 209 of
\cite{Choquet:53}), and its core is defined by the
following:

\vskip8pt\noindent{\bf Definition~A2.5:} the core of $\varphi$
defined above is the set of probability measures that are set-wise
dominated by $\varphi$, i.e. $\mbox{Core}(\varphi) :=
\mbox{Core}(\Gamma,P) = \{Q:\;Q(A)\leq\varphi(A)\;
\mbox{all}\;A\;\mbox{measurable}\}$.\vskip4pt

We add useful regularity properties of Choquet capacity
functionals:

\vskip8pt\noindent{\bf Lemma~A2.1:} If $\varphi$ is a Choquet
capacity functional, by the Choquet Capacitability Theorem
(section 38.2 page 232 of \cite{Choquet:53}), in addition to
properties (i)-(iv) of Definition~A2.1, it satisfies
\begin{itemize} \item[(v)] $\varphi(A) = \sup\{\varphi(F):\;
F\subseteq A,\; F\in{\cal F}\}$ for all $A\in{\cal B}$,
\item[(vi)] $\varphi(A) = \inf\{\varphi(G):\; A\subseteq G,\;
G\in{\cal G}\}$ for all $A\in{\cal B}$.\end{itemize}\vskip4pt

Several notions extend integration in case of non-additive
measures. We only use explicitely the notion of Choquet integral,
which we define below.

\vskip8pt\noindent{\bf Definition~A2.6:} The Choquet integral of a
bounded measurable function $f$ with respect to a capacity
$\varphi$ is defined by
\begin{eqnarray}\int_{\mathrm{Ch}}\,f\,\mbox{d}\varphi = \int_0^{\infty}\varphi(\{f\geq
x\})\,\mbox{d}x + \int_{-\infty}^0\,(\varphi(\{f\geq
x\})-1)\,\mbox{d}x,\label{Choquet
integral}.\end{eqnarray}\vskip4pt

The Choquet integral reduces to the Lebesgue integral when
$\varphi$ is a probability measure. In addition, it has a very
simple expression in case $\varphi$ is a Choquet capacity
functional (see Theorem~1 of \cite{CMM:2004}).

\vskip6pt\noindent{\bf Lemma~A2.2:} If $\varphi$ is a Choquet
capacity functional, then for all $f$ bounded measurable, the
Choquet integral of $f$ with respect to $\varphi$ is given by
$\int_{\mathrm{Ch}}\,f\,\mbox{d}\varphi =
\sup_{Q\in\mathrm{Core}(\varphi)}\int\,f\,\mbox{d}Q$.

\subsection*{Appendix B: Proofs of the results in the main text}

\subsubsection*{Reader's guide to the proofs:}

In the proof of Theorem~1, a result very close to (ii)$\iff$(iv)
is stated in \cite{Wasserman:90}, but the proof is
essentially omitted. The proof of (i)$\iff$(iii) relies on
Corollary~1 of \cite{CMM:2004}, which allows to generalize
Proposition~1 of \cite{Jovanovic:89}. The proof of
(iv)$\iff$(v) is straightforward, whereas the proof of
(iii)$\iff$(v) is similar to Theorem~2. The latter is a simple
application of lemma~1, which itself is a simplification of the
main generalized Monge-Kantorovitch duality theorem of
\cite{Kellerer:84}. Lemma~1[a] is lemma~11.8.5 of
\cite{Dudley:2003}.
The proof given here for
completeness is due to N. Belili. The rest of Theorem~2 is a
specialization of the duality result to zero-one cost, which can
also be proved using Proposition (3.3) page 424 of
\cite{Kellerer:84}, but we give a direct proof to show that
we can specialize to closed sets, a fact that we use in the
discussion of the power of the test.\vskip6pt

Theorem~3a is straightforward. Theorem~3b is structured around the
inequality
\[\sup_{{\cal S}_b}\mathbb{G}_n\leq\sup_{\hat{\cal
S}_{b,h_n}}\mathbb{G}_n\\\leq\sup_{{\cal S}_{b,l_n}}\mathbb{G}_n\]
which holds on an event of large enough probability, with suitable
bandwidth sequences $h_n\ll l_n$. Then, lemma~3a shows that
$\sup_{{\cal S}_{b,l_n}}\mathbb{G}_n$ converges weakly to the same
limit as $\sup_{{\cal S}_{b}}\mathbb{G}_n$, namely $\sup_{{\cal
S}_{b}}\mathbb{G}$. Finally, the same reasoning is invoked to show
that $\sup_{\hat{\cal S}_{b,h_n}}\mathbb{G}$ also converges to the
same limit (but for this we need to assume that the bandwidth
satisfies condition (\ref{bandwidth3}) rather than
(\ref{bandwidth}) and (\ref{bandwidth2})). Lemma~3a relies on the
construction of a local empirical process relative to the thin
sets $A\backslash A_b$, where $A$ is in ${\cal S}_{b,l_n}$ and
$A_b$ is in ${\cal S}_{b}$ and is close to $A$ in terms of
Haussdorf metric (hence the term ``thin'').\vskip6pt

Lemma~3b, like Appendix~A1, brings together some facts that are
scattered in \cite{VW:96}. Theorem~3c uses the regulatiry
properties of Choquet capacity functionals to show that finite
unions of balls are core determining. Given a closed set $F$,
using outer regularity of $P$ and a compactness argument, a
decreasing sequence of finite unions of open balls is constructed
that satisfies two requirements: it converges to $F$ both in
$P$-measure and in Haussdorf distance. The regularity properties
of the correspondence $\Gamma$ are then used to control the
Haussdorf distance between the images by $\Gamma$ of $F$ and the
approximating sequence. The absolute continuity of $\nu$ is then
invoqued to conclude, so that the sign of the inequality is
maintained by continuity.  Theorem~3d ties in the problem of
finding core determining classes with the Monge-Kantorovitch dual
under zero-one cost: pairs $(1_F,-1_{\Gamma(F)})$ with $F$ in the
larger class are shown to be convex combinations of pairs
$(1_A,-1_{\Gamma(A)})$ with $A$ in the potential core determining
class.\vskip6pt

\subsubsection*{Proof of Theorem 1:} [a] We first show equivalences (i)$\iff$(iv)$\iff$(ii):\\
Call $\Delta(B)$ the set of all Borel probability measures with
support $B$. Under Assumption~1, the map
$y\mapsto\Delta(\Gamma(y))$ is a map from ${\cal Y}$ to the set of
all non-empty convex sets of Borel probability measures on ${\cal
U}$ which are closed with respect to the weak topology. Moreover,
for any $f\in C_b({\cal U})$, the set of all continuous bounded
real functions on ${\cal U}$, the map \begin{eqnarray*}
y\longmapsto\sup\left\{\int f d\mu:
\mu\in\Delta(\Gamma(y))\right\}=\max_{u\in\Gamma(y)}
f(u)\end{eqnarray*} is ${\cal B}_{\cal Y}$-measurable, so that, by
Theorem~3 of \cite{Strassen:65}, for a given
$\nu\in\Delta({\cal U})$, there exists $\pi$ satisfying
(\ref{disintegration}) with $\pi(y,.)\in\Delta(\Gamma(y))$ for
$P$-almost all $y$ if and only if \begin{eqnarray}\int_{\cal U}
f(u) \nu(du)\leq\int_{\cal Y}\sup_{u\in\Gamma(y)}f(u)
P(dy)\label{Strassen's condition}\end{eqnarray} for all $f\in
C_b({\cal U})$. Now, defining $\overline{P}$ as the set function
\begin{eqnarray*}\overline{P}:\; B\rightarrow P(\{y\in
{\cal Y}:\;\Gamma(y)\cap B\neq\varnothing\}),\end{eqnarray*}
 the right-hand side of (\ref{Strassen's condition}) is shown in
 the following sequence of equalities to be equal to the integral of $f$ with
 respect to $\overline{P}$ in the sense of Choquet (defined by (\ref{Choquet integral})).
\begin{eqnarray} &&\int_{\cal Y}
\sup_{u\in\Gamma(y)}\left\{f(u)\right\}\,P(dy)\nonumber\\ &=&
\int_0^{\infty}P\bigl\{y\in {\cal Y}:
\sup_{u\in\Gamma(y)}\left\{f(u)\right\}\geq x\bigr\}\,\mbox{d}x +
\int_{-\infty}^0\,(P\bigl\{y\in {\cal Y}: \sup_{u\in\Gamma(y)}
\left\{f(u)\right\}\geq x\bigr\}-1)\,\mbox{d}x\nonumber\\
&=&\int_0^{\infty}P\bigl\{y\in {\cal Y}:\,
\Gamma(y)\subseteq\left\{f\geq x\right\}\bigr\}\,\mbox{d}x +
\int_{-\infty}^0\,(P\bigl\{y\in {\cal Y}:\,
\Gamma(y)\subseteq\left\{f\geq
x\right\}\bigr\}-1)\,\mbox{d}x\nonumber\\&=&\int_0^{\infty}\overline{P}(\{f\geq
x\})\,\mbox{d}x + \int_{-\infty}^0\,(\overline{P}(\{f\geq
x\})-1)\,\mbox{d}x=\int_{\mathrm{Ch}}\,f\,\mbox{d}\overline{P}.\nonumber\end{eqnarray}

By Theorem~1 of \cite{CMM:2004}, for any $f\in C_b({\cal
U})$,
\begin{eqnarray*} \int_{\mathrm{Ch}}\,f\,\mbox{d}\overline{P}
=\max_{\gamma\in{\mathrm Sel}(\Gamma)}\int_{\cal U} f(u)
P\gamma^{-1}(du),\end{eqnarray*} so that (\ref{Strassen's
condition}) is equivalent to
\begin{eqnarray} \max_{\gamma\in{\mathrm Sel}(\Gamma)}\int_{\cal U} f(u)
P\gamma^{-1}(du)\geq\int_{\cal U} f(u) \nu(du)\label{Strassen
Choquet}\end{eqnarray} for any $f\in C_b({\cal U})$. If $\nu$ is
in the weak closure of the set of convex combinations of elements
of $\{P\gamma^{-1}: \gamma\in{\mathrm{Sel}}(\Gamma)\}$, then by
linearity of the integral and the definition of weak convergence,
(\ref{Strassen Choquet}) holds. Conversely, if $\nu$ satisfies
(\ref{Strassen Choquet}), then it satisfies
\begin{eqnarray*}
\int_{\mathrm{Ch}}\,f\,\mbox{d}\overline{P}\geq\int_{\cal U} f(u)
\nu(du)\end{eqnarray*} and by monotone continuity, we have for all
$A\in{\cal B}_{\cal U}$, and $1_A$ the indicator function,
\begin{eqnarray*}\int_{\cal U} 1_{A}(u)
\nu(du)\leq\int_{{\mathrm Ch}}1_{A} d\overline{P}.\end{eqnarray*}
Hence $\nu(A)\leq\overline{P}(A)$ for all $A\in{\cal B}_{\cal U}$,
which by Corollary~1 of \cite{CMM:2004} implies that $\nu$
is the weak limit of a sequence of convex combinations of elements
of $\{P\gamma^{-1}: \gamma\in{\mathrm{Sel}}(\Gamma)\}$, hence it
is a mixture in the desired sense and the proof is
complete.\vskip6pt

[b] We now show equivalences (iii)$\iff$(iv)$\iff$(v): \\ Using
theorem~2 below, it suffices to show that (\ref{plausibility
constraint}) is equivalent to $\nu(\Gamma(A))\geq P(A)$ for all
$A\in{\cal B}_{\cal Y}$. As previously, define $\overline{P}$ as
the set function on ${\cal B}_{\cal U}$
\begin{eqnarray*}\overline{P}:\; B\rightarrow P(\{y\in
{\cal Y}:\;\Gamma(y)\cap B\neq\varnothing\}).\end{eqnarray*}
Define also $\underline{P}$ as the set function
\begin{eqnarray*}\underline{P}:\; B\rightarrow P(\{y\in
{\cal Y}:\;\Gamma(y)\subseteq B\}).\end{eqnarray*} Since
$\overline{P}(B)=1-\underline{P}(B^c)$, we have the well known
equivalence between $\nu(B)\leq\overline{P}(B)$ for all $B\in{\cal
B}_{\cal U}$ and $\nu(B)\geq\underline{P}(B)$ for all $B\in{\cal
B}_{\cal U}$. In particular, for $B=\Gamma(A)$ for any $A\in{\cal
B}_{\cal Y}$, we have $\nu(B)\subseteq\{y\in{\cal
Y}:\,\Gamma(y)\subseteq\Gamma(A)\}$. As $A\subseteq\{y\in{\cal
Y}:\,\Gamma(y)\subseteq\Gamma(A)\}$, we have $\nu(\Gamma(A))\geq
P(B)$. Conversely, for some $B\in{\cal B}_{\cal U}$, call
$B_*=\{y\in{\cal Y}:\,\Gamma(y)\subseteq B\}$. Then, we have
$P(B_*)\leq\nu(\Gamma(B_*))$. The result follows from the
observation that $\Gamma(B_*)\subseteq B$.

\subsubsection*{Proof of Theorem~1':} The proof completely
parallels the proof of Theorem~1. The equivalence between 1(iii)
and 1'(iii') drives the equivalence of each of the formulations in
Theorem~1' with each of the formulations in Theorem~1.

\subsubsection*{Lemma 1:} If
$\varphi:{\cal Y}\times{\cal U}\rightarrow\mathbb{R}$ is bounded,
non-negative and lower semicontinuous, then
\begin{eqnarray*}\inf_{\pi\in{\cal M}(P,\nu)}\;\pi\varphi
=\sup_{f\oplus g\leq\varphi}\;(Pf+\nu g)\end{eqnarray*}

\subsubsection*{Proof of Lemma 1:} It can be shown to be a special case of
corollary (2.18) of \cite{Kellerer:84}; however, a direct
proof is more transparent, so we give it here for completeness.
The left-hand side is immediately seen to be always larger than
the right-hand side, so we show the reverse inequality. \vskip6pt
[a] case where $\varphi$ is continuous and ${\cal U}$ and ${\cal
Y}$ are compact.\\Call $G$ the set of functions on ${\cal
Y}\times{\cal U}$ strictly dominated by $\varphi$ and call $H$ the
set of functions of the form $f+g$ with $f$ and $g$ continuous
functions on ${\cal Y}$ and ${\cal U}$ respectively. Call
$s(c)=Pf+\nu g$ for $c\in H$. It is a well defined linear
functional, and is not identically zero on $H$. $G$ is convex and
sup-norm open. Since $\varphi$ is continuous on the compact ${\cal
Y}\times{\cal U}$, we have
\begin{eqnarray*}s(c)\leq\sup f+\sup g<\sup\varphi\end{eqnarray*} for
all $c\in G\cap H$, which is non empty and convex. Hence, by the
Hahn-Banach theorem, there exists a linear functional $\eta$ that
extends $s$ on the space of continuous functions such that
\begin{eqnarray*}\sup_{G}\;\eta=\sup_{G\cap H}\;s.\end{eqnarray*}
By the Riesz representation theorem, there exists a unique finite
non-negative measure $\pi$ on ${\cal Y}\times{\cal U}$ such that
$\eta(c)=\pi c$ for all continuous $c$. Since $\eta=s$ on $H$, we
have \begin{eqnarray*}\int_{{\cal Y}\times{\cal U}}
f(y)\;d\pi(y,u)&=&\int_{\cal Y}f(y)\;dP(y)\\
\int_{{\cal Y}\times{\cal U}} g(u)\;d\pi(y,u)&=&\int_{\cal
Y}g(u)\;d\nu(y),\end{eqnarray*} so that $\pi\in{\cal M}(P,\nu)$
and \begin{eqnarray*}\sup_{f\oplus g\leq\varphi} (Pf+\nu
g)=\sup_{G\cap H} s=\sup_{H} \eta=\pi\varphi.\end{eqnarray*}
\vskip8pt [b] ${\cal Y}$ and ${\cal U}$ are not necessarily
compact, and $\varphi$ is continuous.\vskip6pt For all $n>0$,
there exists compact sets $K_n$ and $L_n$ such that
\begin{eqnarray*} \max\left(P({\cal Y}\backslash K_n),\nu({\cal
U}\backslash L_n)\right)\leq\frac{1}{n}.\end{eqnarray*} Let
$(a,b)$ be an element of ${\cal Y}\times{\cal U}$ and define two
probability measures $\mu_n$ and $\nu_n$ with compact support by
\begin{eqnarray*}\mu_n(A)&=&P(A\cap K_n)+P(A\backslash K_n)\delta_a(A)\\
\nu_n(B)&=&\nu(B\cap L_n)+\nu(B\backslash
L_n)\delta_b(B),\end{eqnarray*} where $\delta$ denotes the Dirac
measure. By [a] above, there exists $\pi_n$ with marginals $\mu_n$
and $\nu_n$ such that
\begin{eqnarray*}\pi_n\varphi\leq\sup_{f\oplus g\leq\varphi}
(Pf+\nu g)+\frac{\varphi(a,b)}{n}.\end{eqnarray*} Since $(\pi_n)$
has weakly converging marginals, it is weakly relatively compact.
Hence it contains a weakly converging subsequence with limit
$\pi\in{\cal M}(P,\nu)$. By Skorohod's almost sure representation
(see for instance theorem 11.7.2 page 415 of
\cite{Dudley:2003}), there exists a sequence of random
variables $X_n$ on a probability space $(\Omega,{\cal
A},\mathbb{P})$ with law $\pi_n$ and a random variable $X_0$ on
the same probability space with law $\pi$ such that $X_0$ is the
almost sure limit of $(X_n)$. By Fatou's lemma, we then have
\begin{eqnarray*}\mbox{liminf}\;\pi_n\varphi
= \mbox{liminf}\, \mathbb{E} \varphi(X_n) \geq \mathbb{E}
\,\mbox{liminf} \varphi(X_n) = \mathbb{E} \varphi(X_0) =
\pi\varphi.\end{eqnarray*} Hence we have the desired result.
\vskip8pt [c] General case.\vskip6pt $\varphi$ is the pointwise
supremum of a sequence of continuous bounded functions, so the
result follows from upward $\sigma$-continuity of both
$\inf_{\pi\in{\cal M}(P,\nu)}\pi\varphi$ and $\sup_{f\oplus
g\leq\varphi} (Pf+\nu g)$ on the space of lower semicontinuous
functions, shown in propositions (1.21) and (1.28) of
\cite{Kellerer:84}.


\subsubsection*{Proof of Theorem 2:} Under assumption~1,
$\Gamma$ is closed valued, hence
$\varphi(y,u)=1_{\{u\notin\Gamma(y)\}}$ is lower semicontinuous
and (\ref{dual Monge-Kantorovich with functions}) is a direct
application of lemma~1 above.

We now show (\ref{dual Monge-Kantorovich with sets}). Since the
sup-norm of the cost function is 1 (the cost function is an
indicator), the supremum in (\ref{dual Monge-Kantorovich with
functions}) is attained pairs of functions $(f,g)$ in ${\cal F}$,
defined by \begin{eqnarray*}{\cal
F}=\{(f,g)\in\mathbb{L}^1(P)\times\mathbb{L}^1(\nu),\;0\leq
f\leq1,\;-1\leq
g\leq0,\\f(y)+g(u)\leq1_{\{u\notin\Gamma(y)\}},\;\mbox{$f$ upper
semicontinuous}\}.\end{eqnarray*} Now, $(f,g)$ can be written as a
convex combination of pairs $(1_A,-1_B)$ in ${\cal F}$. Indeed,
$f=\int_{0}^{1}1_{\{f\geq x\}}\,dx$ and $g=\int_{0}^{1}-1_{\{g\leq
-x\}}\,dx$, and for all $x$, $1_{\{f\geq
x\}}(y)-1_{\{g\leq-x\}}(u)\leq1_{\{u\notin\Gamma(y)\}}$. Since the
functional on the right-hand side of (\ref{dual Monge-Kantorovich
with functions}) is linear, the supremum is attained on such a
pair $(1_A,-1_B)$. Hence, the right-and side of (\ref{dual
Monge-Kantorovich with functions}) specializes to
\begin{eqnarray}\sup_{A\times B\subseteq D}
(P(A)-1+\nu(B)).\label{duality with indicators}\end{eqnarray} For
$D=\{(y,u):\, u\notin\Gamma(y)\}$, $A\times B\subseteq D$ means
that if $y\in A$ and $u\in B$, then $u\notin\Gamma(y)$. In other
words $u\in B$ implies $u\notin\Gamma(A)$, which can be written
$B\subseteq\Gamma(A)^{c}$. Hence, the dual problem can be written
\begin{eqnarray*}\sup_{\Gamma(A)\subseteq B^c}
(P(A)-1+\nu(B))=\sup_{\Gamma(A)\subseteq B}
(P(A)-\nu(B)).\end{eqnarray*} and (\ref{dual Monge-Kantorovich
with sets}) follows immediately.\vskip6pt


\subsubsection*{Proof of Theorem~3a:}
Let $A_0$ be the subset of ${\cal Y}$ that achieves the maximum of
$\delta(A)=P(A)-\nu(\Gamma(A))$ over $A\in{\cal S}\backslash{\cal
S}_b$. Call $\delta_0=\delta(A_0)$, and note that $\delta_0<0$. We
have \begin{eqnarray*} \sqrt{n}T_{2^{\cal Y}}(P_n,\Gamma,\nu) &=&
\sup_{A\in2^{\cal Y}} [\mathbb{G}_n(A) + \sqrt{n}(P(A) +
\nu(\Gamma(A)))]\\ &=& \max\{ \sup_{{\cal S}_b} \mathbb{G}_n ,
\sup_{A\in2^{\cal Y}\backslash{\cal S}_b} [\mathbb{G}_n(A) +
\sqrt{n}(P(A) + \nu(\Gamma(A)))] \}. \end{eqnarray*} The second
term in the maximum of the preceding display is dominated by
\[\sup_{2^{\cal Y}\backslash{\cal S}_b} \mathbb{G}_n + \sqrt{n}
\delta_0,\]whose limsup is almost surely non-positive. Hence
(\ref{asymptotic distribution}) follows from the convergence of
the empirical process. (\ref{asymptotic approximation}) follows
from the fact that, under (\ref{bandwidth}), for all $n$
sufficiently large, $\hat{\cal S}_{b,h_n}$ is almost surely equal
to ${\cal S}_b$.

\subsubsection*{Proof of Theorem~3b:}
Consider two sequences of positive numbers $l_n$ and $h_n$ such
that they both satisfy (\ref{bandwidth3}), $l_n>h_n$ and
$(l_n-h_n)^{-1}\sqrt{\frac{\ln\ln n}{n}}\rightarrow0$. Notice that
$\{\varnothing,{\cal Y}\} \subseteq {\cal S}_b,{\cal
S}_{b,h},\hat{\cal S}_{b,h}$ for any $h>0$. Since
$\mathbb{G}_n({\cal Y})=0$, we therefore have $\sup_{{\cal
S}_b}\mathbb{G}_n$, $\sup_{{\cal S}_{b,l_n}}\mathbb{G}_n$ and
$\sup_{\hat{\cal S}_{b,h_n}}\mathbb{G}_n$ non-negative. Hence,
calling $\zeta_n$ the indicator function of the event $\sup_{\cal
S}\mathbb{G}_n\leq(l_n-h_n)\sqrt{n}$, we can write
\begin{eqnarray*}\zeta_n \sup_{{\cal S}_b}\mathbb{G}_n&\leq&
\zeta_n \max\left\{\sup_{{\cal
S}_b}[\mathbb{G}_n+\sqrt{n}(P-\nu\Gamma)], \sup_{{\cal
S}\backslash{\cal
S}_b}[\mathbb{G}_n+\sqrt{n}(P-\nu\Gamma)]\right\}\\
&\leq&\zeta_n \sqrt{n} T_{\cal S}(P_n,\Gamma,\nu)\\
&\leq&\zeta_n \sup_{\hat{\cal
S}_{b,h_n}}\mathbb{G}_n\\&\leq&\zeta_n\sup_{{\cal
S}_{b,l_n}}\mathbb{G}_n,\end{eqnarray*} where the first inequality
holds because the left-hand side is equal to the first term in the
right-hand side, the second inequality holds trivially as an
equality since ${\cal S}={\cal S}_b\cup{\cal S}\backslash{\cal
S}_b$, the third inequality holds because on ${\cal
S}\backslash\hat{\cal S}_{b,h_n}$, we have by definition
$\mathbb{G}_n + \sqrt{n}(P-\nu\Gamma) = \sqrt{n}(P_n-\nu\Gamma)
\leq -h_n \leq 0$, and the last inequality holds because on
$\{\zeta_n=1\}$, we have that $A\in\hat{\cal S}_{b,h_n}$ implies
$\nu\Gamma(A) \leq P_n(A) + h_n = P(A) + (P_n-P)(A) +h_n \leq P(A)
+ \sup_{\cal S}\mathbb{G}_n/\sqrt{n} + h_n \leq P(A) + l_n - h_n +
h_n = P(A) + l_n $, which implies that $A\in{\cal
S}_{b,l_n}$.\vskip6pt

By Lemma~3a and Appendix~A1, we have that both $\sup_{{\cal
S}_b}\mathbb{G}_n$ and $\sup_{{\cal S}_{b,l_n}}\mathbb{G}_n$
converge weakly to $\sup_{{\cal S}_b}\mathbb{G}$. It is shown
below that $\zeta_n\rightarrow_p1$, so that Slutsky's lemma (lemma
2.8 page 11 of \cite{Vaart:98}) yields the weak convergence
of $\zeta_n\sup_{{\cal S}_b}\mathbb{G}_n$ and $\zeta_n\sup_{{\cal
S}_{b,l_n}}\mathbb{G}_n$ to the same limit, and hence that of
$\zeta_n T_{\cal S}(P_n,\Gamma,\nu)$ and $\zeta_n\sup_{\hat{\cal
S}_{b,h_n}}\mathbb{G}_n$. It follows from Slutsky's lemma again
that
\begin{eqnarray*}\sqrt{n} T_{\cal S}(P_n,\Gamma,\nu)\rightsquigarrow
\sup_{{\cal S}_b}
\mathbb{G}\hskip10pt\mbox{and}\hskip10pt\sup_{\hat{\cal
S}_{b,h_n}} \mathbb{G}_n\rightsquigarrow \sup_{{\cal S}_b}
\mathbb{G},\end{eqnarray*} which proves (\ref{asymptotic
distribution}).\vskip6pt We now prove that
$\zeta_n\rightarrow_p1$. Indeed, for any $\epsilon>0$, $P(
|\zeta_n-1| > \epsilon ) = P( \zeta_n = 0 ) = P( \sup_{\cal
S}\mathbb{G}_n > (l_n-h_n)\sqrt{n} ) \rightarrow 0$ by the Law of
Iterated Logarithm, since $(l_n-h_n)\sqrt{n}\gg\sqrt{\ln\ln n}$ by
assumption.\vskip6pt

There remains to show (\ref{asymptotic approximation}). Defining
$\xi_n$ as the indicator of the set \[ \{-h_n \sqrt{n} \leq
\sup_{\cal S}\mathbb{G}_n \leq (l_n-h_n) \sqrt{n}\},\] we have the
inequalities \[ \xi_n \sup_{{\cal S}_{b}} \mathbb{G} \leq \xi_n
\sup_{{\hat{\cal S}}_{b,h_n}} \mathbb{G} \leq \xi_n \sup_{{\cal
S}_{b,l_n}} \mathbb{G}.\] Indeed, the first inequality holds
because $\sup_{\cal S} \mathbb{G}_n \geq -h_n \sqrt{n}$ implies
that $P_n(A) \geq P(A) - h_n$ for all $A$, hence that ${\cal S}_b
\subseteq \hat{\cal S}_{b,h_n}$; and the second inequality holds
because because on $\{\xi_n=1\}$, we have that $A\in\hat{\cal
S}_{b,h_n}$ implies $\nu\Gamma(A) \leq P_n(A) + h_n = P(A) +
(P_n-P)(A) +h_n \leq P(A) + \sup_{\cal S}\mathbb{G}_n/\sqrt{n} +
h_n \leq P(A) + l_n - h_n + h_n = P(A) + l_n $, which implies that
$A\in{\cal S}_{b,l_n}$.\vskip6pt

By Lemma~3a suitably modified to apply to the oscillations of
$\mathbb{G}$ instead of the oscillations of $\mathbb{G}_n$, we
have that $\sup_{{\cal S}_{b,l_n}}\mathbb{G}$ converges weakly to
$\sup_{{\cal S}_b}\mathbb{G}$. It is shown below that
$\xi_n\rightarrow_p1$, so that Slutsky's lemma  yields the weak
convergence of $\xi_n\sup_{{\cal S}_b}\mathbb{G}_n$ and
$\xi_n\sup_{{\cal S}_{b,l_n}}\mathbb{G}$ to the same limit, and
hence that of $\xi_n\sup_{\hat{\cal S}_{b,h_n}}\mathbb{G}$. It
follows from Slutsky's lemma again that
\begin{eqnarray*}\sup_{\hat{\cal
S}_{b,h_n}} \mathbb{G}\rightsquigarrow \sup_{{\cal S}_b}
\mathbb{G},\end{eqnarray*} which proves (\ref{asymptotic
approximation}).\vskip6pt We now prove that $\xi_n\rightarrow_p1$.
Indeed, for any $\epsilon>0$, $P( |\xi_n-1| > \epsilon ) = P(
\zeta_n = 0 ) = P( \sup_{\cal S}\mathbb{G}_n > (l_n-h_n)\sqrt{n}
\; \mbox{or} \sup_{\cal S}\mathbb{G}_n < -h_n \sqrt{n} )
\rightarrow 0$ by the Law of Iterated Logarithm, since
$(l_n-h_n)\sqrt{n}\gg\sqrt{\ln\ln n}$ and $h_n
\sqrt{n}\gg\sqrt{\ln\ln n}$ by assumption.\vskip6pt

\subsubsection*{Proof of Lemma~3a:}
Take a bandwidth sequence $l_n$ that satisfies (\ref{bandwidth3}),
and take ${\cal S}_{b,l_n}$ as in definition~3.3. Under
assumption~FS, take $A\in {\cal S}_{b,l_n}$ and an $A_{0}\in{\cal
S}_b$ such that $d_H\left( A,A_{0}\right) \leq
\zeta_n=Kl_{n}^{\eta}$ (we suppress the dependence of $A_b$ on $A$
for ease of notation). As ${\cal S}_{b}\subseteq{\cal S}_{b,l_n}$,
one has
\begin{equation}
\sup_{A\in {\cal S}_b}\mathbb{G}_n(A)\leq \sup_{B\in{\cal
S}_{b,l_n}}\mathbb{G}_{n}(A)  \label{one}
\end{equation}
Second, since $A_b\subseteq A$, one has
\begin{eqnarray*}\sup_{A\in
\mathcal{S}_{b,l_{n}}}\mathbb{G}_{n}(A)&=&\sup_{A\in \mathcal{S}
_{b,l_{n}}}\left[ \mathbb{G}_{n} (A_b) +
\mathbb{G}_{n}(A\backslash A_b) \right]\\&\leq& \sup_{A\in
\mathcal{S} _{b,l_{n}}}\left[ \mathbb{G}_{n} (A_b)\right] +
\sup_{A\in \mathcal{S} _{b,l_{n}}}\left[\mathbb{G}_{n}(A\backslash
A_b) \right].\end{eqnarray*}  If we have that
\begin{eqnarray*}
\sup_{A\in{\cal S}_{b,l_n}}\left\vert \mathbb{G}_{n}(A\backslash
A_b) \right\vert = O_{\mathrm{a.s.}}\left( \sqrt{ \zeta_{n}\ln\ln
n }\right),
\end{eqnarray*}
then
\begin{equation}
\sup_{A\in \mathcal{S}_{b,l_{n}}}\mathbb{G}_{n}(A)=\sup_{A\in
\mathcal{S}_{b,l_{n}}}\left[ \mathbb{G}_{n}(A_b) \right] +
O_{\mathrm{a.s.}}\left( \sqrt{\zeta_{n}\ln\ln n}\right)
\label{two}
\end{equation} noting the dependence of $A_b$ on $A$ in the
expression above. But since $A_b \in \mathcal{S}_{b}$, one has
$\sup_{A\in \mathcal{S}_{b,l_{n}}}\left[ \mathbb{G}_{n}\left(
A_b\right) \right] \leq \sup_{A\in
\mathcal{S}_{b}}\mathbb{G}_{n}(A)$. This fact, along with
(\ref{one}) and (\ref{two}), yields the result.\vskip6pt

We now show that we have indeed that \begin{eqnarray*}
\sup_{A\in{\cal S}_{b,l_n}}\left\vert \mathbb{G}_{n}(A\backslash
A_b) \right\vert = O_{\mathrm{a.s.}}\left( \sqrt{ \zeta_{n}\ln\ln
n }\right). \end{eqnarray*}

This relies on the construction of a local empirical process
relative to the thin regions $A\backslash A_b$. First consider
such a region. If $A\in{\cal S}_{b}$, the result holds trivially,
so that we may assume that $A\in{\cal S}_{b,l_n}\backslash{\cal
S}_b$, so that $A\backslash A_b$ is not empty. We distinguish the
case where $A$ is a bounded rectangle, and the cases where $A$ is
unbounded.\vskip6pt

\begin{itemize}

\item[(i)] $A$ is a bounded rectangle, i.e. of the form
$(y_1,z_1)$ $\times$ $\ldots$ $\times$ $(y_{d_y},z_{d_y})$, with
$y_1,$ $\ldots,$ $y_{d_y},z_1,$ $\ldots,$ $z_{d_y}$ real. Then,
since $d_{H}(A,A_b)\leq \zeta_n$, $A_b$ is also a bounded
rectangle, and the $A\backslash A_b$ is the union of at least one
(since $A$ and $A_b$ are distinct) and at most $f(d_y)$ (the
number of faces of a rectangle in $\mathbb{R}^{d_y}$) rectangles
with at least one dimension bounded by $\zeta_n$.

\item[(ii)] $A$ is an unbounded rectangle, i.e. of the same form
as above, except that some of the edges are $+\infty$ of
$-\infty$. Then $A_b$ is also an unbounded rectangle, and
$A\backslash A_b$ is also the union of a finite number of
rectangles with one dimension bounded by $\zeta_n$.

\end{itemize}
In both cases $(i)$, and $(ii)$, $A\backslash A_b$ is the union of
a finite number of rectangles with at least one dimension bounded
by $\zeta_n$. Hence if we control the supremum of the empirical
process on one of these thin rectangles, when $A$ ranges over
${\cal S}_{b,l_n}$, we can control it on $A\backslash
A_b$.\vskip6pt

Hence, it suffices to prove that \begin{eqnarray*} \sup_{A\in{\cal
S}_{b,l_n}}\left\vert \mathbb{G}_{n}(\varphi_n (A)) \right\vert =
O_{\mathrm{a.s.}}\left( \sqrt{ \zeta_n\ln\ln n }\right),
\end{eqnarray*} where $\varphi_n$ is the homothety that carries
$A$ into one of the thin rectangles described above.\vskip6pt

As an homothety, $\varphi_n$ is invertible and bi-measurable, and
since $\varphi_n(A)$ has at least one dimension bounded by
$\zeta_n$, and $P$ is absolutely continuous with respect to
Lebesgue measure, $P(\varphi_n(A))=O(\zeta_n)$ uniformely when $A$
ranges over ${\cal S}_{b,l_n}$. Now, for any $A\in{\cal
S}_{b,l_n}$, we have
\begin{eqnarray*}\mathbb{G}_{n}(\varphi_n
(A))&=&\sqrt{n}\left[P_n(\varphi_n(A))-P(\varphi_n(A))\right]\\&=&\frac{1}{\sqrt{n}}
\sum_{i=1}^{n}\left(
1_{\{\varphi_n(A)\}}(Y_i)-\mathbb{E}_{P}(1_{\{\varphi_n(A)\}}(Y))\right)\\
&=& \frac{1}{\sqrt{n}} \sum_{i=1}^{n}\left(
1_{A}(\varphi_n^{-1}(Y_i))-\mathbb{E}_{P}(1_{A}(\varphi_n^{-1}(Y)))\right)\\
&:=&\sqrt{\zeta_n} L_n(1_A,\varphi_n),\end{eqnarray*}

where $L_n(1_A,\varphi_n)$ is defined as
\begin{eqnarray*}\frac{1}{\sqrt{n \zeta_n}} \sum_{i=1}^{n}\left(
1_{A}(\varphi_n^{-1}(Y_i))-\mathbb{E}_{P}(1_{A}(\varphi_n^{-1}(Y)))\right)\end{eqnarray*}
to conform with the notation of \cite{EM:97}.\vskip6pt

Conditions~A(i)-A(iv) of the latter hold for $a_n=b_n=l_n$ and
$a=0$ under (\ref{bandwidth3}), and conditions~S(i)-S(iii) and
F(ii) and F(iv)-F(viii) hold because ${\cal F}$ is here the class
of indicator functions of ${\cal S}_{b,l_n}$ which, as a subclass
of ${\cal S}$, is a Vapnik-$\breve{\mbox{C}}$ervonenkis class of
sets. Hence Theorem~1.2 of \cite{EM:97} holds, and
\begin{eqnarray*}\sup_{A\in{\cal S}_{b,l_n}}\left\vert L_n(1_A,\varphi_n)\right\vert
= O_{\mathrm{a.s.}}\left( \sqrt{ \ln\ln n }\right)\end{eqnarray*}
so that the desired result holds.

\subsubsection*{Proof of Lemma~3b:}
Consider ${\cal S}=\{\,(y,z):\;(y,z)\in\mathbb{R}^{2d_{y}}\}$. It
is a Vapnik-$\breve{\mbox{C}}$ervonenkis class. Indeed, if
$d_y=1$, its Vapnik-$\breve{\mbox{C}}$ervonenkis index is three,
since ${\cal S}$ can pick out the two elements of a set of
cardinality 2, but can never pick out the subset $\{x,z\}$ of a
set of three elements $\{x,y,z\}$. More generally, it can be shown
that the Vapnik-$\breve{\mbox{C}}$ervonenkis index of ${\cal S}$
is $2d_y+1$ (see Example 2.6.1 page 135 of \cite{VW:96}).
Hence the class ${\cal S}_K$ is also
Vapnik-$\breve{\mbox{C}}$ervonenkis. The latter follows from lemma
2.6.17(iii) page 147 of \cite{VW:96} and the fact that it is
contained in the $K$-iterated union ${\cal
S}\sqcup\ldots\sqcup{\cal S}$, where the ``square union'' of two
classes of sets ${\cal S}_1$ and ${\cal S}_2$ is defined by ${\cal
S}_1\sqcup{\cal S}_2=\{A_1\cup A_2:\;A_1\in{\cal
S}_1,\,A_2\in{\cal S}_2\}$.

\subsubsection*{Proof of Theorem~3c:}
From Fact~2, we know that we can restrict attention to closed
subsets of ${\cal Y}$. Take $F$ one such subset. By the outer
regularity of Borel probability measures, for all $n$ there is an
open set ${\cal O}_n'$ such that $F\subseteq{\cal O}_n'$ and
$P({\cal O}_n')\leq P(F) + 1/n$. Since ${\cal O}_n'$ is open, for
each $y\in F$, there exists $r_y>0$ such that the open ball
$B(y,r_y)$ centered at $y$ with radius $r_y$ is included in ${\cal
O}_n'$, and by construction, the open set $\tilde{\cal
O}_n'=\bigcup_{y\in F} B(y,\mbox{min}(r_y,1/n^2))$ covers $F$. As
a closed subset of a compact set, $F$ is compact. Hence we can
call ${\cal O}_n$ the finite sub-covering of $F$ extracted from
$\tilde{\cal O}_n'$. ${\cal O}_n$ is therefore a finite union of
open balls with positive radii, i.e. it belongs to
$\tilde{S}_{\mathrm{SW}}$. By construction of ${\cal O}_n$, we
have $d_{\mathrm{H}}({\cal O}_n,F)\leq1/n^2$, and we know that
$\Gamma(F)\subseteq\Gamma({\cal O}_n)$, and we shall now show that
$\nu(\Gamma({\cal O}_n))$ converges to $\nu(\Gamma(F))$ to yield
the result that $\tilde{\cal S}_{\mathrm{SW}}$ is core
determining.\vskip4pt

Consider the following partition ${\cal Y}={\cal Y}_I\cup{\cal
Y}_n^-\cup{\cal Y}_n^+$ with:
\begin{eqnarray*}{\cal Y}_I&=&\{y\in{\cal Y}:\;\nu(\Gamma(y))=0\},\\
{\cal Y}_n^-&=&\{y\in{\cal Y}:\;0<\nu(\Gamma(y))<1/n\},\\
{\cal Y}_n^+&=&\{y\in{\cal
Y}:\;\nu(\Gamma(y))\geq1/n\}.\end{eqnarray*}

Define $F_I=F\cap{\cal Y}_I$, $F_n^-=F\cap{\cal Y}_n^-$ and
$F_n^+=F\cap{\cal Y}_n^+$, and similarly for ${\cal O}_n$, with
${\cal O}_n^I$ denoting ${\cal O}_n\cap{\cal Y}_I$.\vskip6pt

Consider first ${\cal O}_n^I\backslash F_I$. Assumption~(CD3)
yields immediately that $\nu(\Gamma({\cal O}_n^I\backslash
F_I))\downarrow0$.\vskip6pt

Consider now ${\cal O}_n^-\backslash F_n^-$. Under
assumption~(CD6), $\nu(\Gamma({\cal Y}_n^-))\downarrow0$, hence
$\nu(\Gamma({\cal O}_n^-\backslash F_n^-))\downarrow0$.\vskip6pt

Consider now ${\cal O}_n^+\backslash F_n^+$. Consider the disjoint
connected components of $\Gamma({\cal O}_n^+)$. Their $\nu$
measure is at least $1/n$ by construction, hence by the
compactness of ${\cal U}$, the number $J_n$ of disjoint connected
components of $\Gamma({\cal O}_n^+)$ is no greater than $n$. We
have shown above that $d_H({\cal O}_n,F)<1/n^2$, hence we have
$d_H({\cal O}_n^+,F_n^+)<1/n^2$. By assumption~(CD5), this implies
that $d_H(\Gamma({\cal O}_n^+),\Gamma(F_n^+))=O(1/n^2)$. Hence for
$n$ sufficiently large, all the disjoint connected components of
$\Gamma({\cal O}_n^+)$ intersect $\Gamma(F_n^+)$. Call
$(C_j)_{j=1}^{J_n}$ the disjoint connected components of
$\Gamma({\cal O}_n^+)$. We have \[\nu(\Gamma({\cal O}_n^+)) =
\sum_{j=1}^{J_n} \nu(\Gamma(C_j))=\sum_{j=1}^{J_n}
\left(\nu(\Gamma(C_j))+O(1/n^2)=\nu(\Gamma(F_n^+))+O(1/n)\right),\]
where the second equality holds under assumption~(CD2). Since
$F_n^+\subseteq{\cal O}_n^+$, we therefore have the desired result
$\nu(\Gamma({\cal O}_n^+\backslash F_n^+))\downarrow0$, which
completes the proof.

\subsubsection*{Proof of Theorem~3d:}
From fact~2, we can restrict attention to closed subsets of ${\cal
Y}=\mathbb{R}$. Call ${\cal Y}_I$ the subset of ${\cal Y}$ defined
by $u(y)=l(y)$ $P$-almost surely (and therefore everywhere since
$u$ and $l$ are increasing). Note that the restriction of
$\nu\Gamma$ to ${\cal Y}_I$ is a probability measure. Consider a
closed subset $F$ of $\cal Y$. Call $F_I=F\cap{\cal Y}_I$ (resp.
$F_U=F\backslash F_I$) the intersection of $F$ with ${\cal Y}_I$
(resp. its complementary). Because of the monotonicity of the
envelopes, $\nu(\Gamma(F))=\nu(\Gamma(F_I))+\nu(\Gamma(F_U))$,
hence we only need to prove the result for closed subsets of
${\cal Y}_I$ and for closed subsets of ${\cal Y}\backslash{\cal
Y}_I$.\vskip6pt

Take $F$ a subset of ${\cal Y}_I$. The restriction
$\nu\Gamma_{|{\cal Y}_I}$ of $\nu\Gamma$ to ${\cal Y}_I$ is a
probability measure, and the class of sets ${\cal C}_I$ defined by
${\cal C}_I=\{A\in{\cal Y}:\;A=\tilde{A}\cap{\cal
Y}_I,\;\tilde{A}\in{\cal C}\}$ is value determining for
$\nu\Gamma_{|{\cal Y}_I}$. By the monotonicity of the envelopes,
we have $\nu(\Gamma(\tilde{A}))=\nu(\Gamma(A)) +
\nu(\Gamma(\tilde{A}\backslash A))$ (with the notation of the
definition of ${\cal C}_I$ above). Hence, if $\nu(\Gamma(A))\geq
P(A)$ for all $A\in{\cal C}$, then $\nu(\Gamma(A))\geq P(A)$ for
all $A\subseteq{\cal Y}_I$.

We can now restrict attention to the case where the upper and
lower envelopes are distinct, in which case, for a closed set $F$,
$\Gamma(F)$ has at most a countable number of connected parts,
which we denote $C_n$, $n\in\mathbb{Z}$, ordered in the sense that
$\inf C_n>\sup C_{n-1}$. By construction, each $C_n$ is the image
by $\Gamma$ of a subset $F_n$ of $F$. $\Gamma$ being
convex-valued, the monotonicity of the envelopes $u$ and $l$
implies upper-semicontinuity of $l$ and lower-semicontinuity of
$u$. Therefore, $C_n=\Gamma(F_n)=\Gamma([\inf F_n,\sup F_n])$, and
we deduce that $\nu\Gamma(F)=\nu\Gamma(\bigcup_{n}I_n)$ where
$(I_n)_{n\in\mathbb{Z}}$ is a countable collection of disjoint
closed intervals in $\mathbb{R}$. Hence if we show that
$\nu\Gamma(I) \geq P(I)$ for any interval $I$, then we have
$\nu\Gamma(F) = \sum_{n}\nu\Gamma(I_n) \geq \sum_{n}P(I_n) \geq
P(F)$, and the inequality holds for $F$.\vskip4pt

Now, for any $y_1<y_2\in\mathbb{R}$ we have $P(y_1,y_2] =
P(y_1,+\infty) + P(-\infty,y_2] - 1 \leq \nu\Gamma(y_1,+\infty) +
\nu\Gamma(-\infty,y_2] - 1 = \nu(u(y_2) - l(y_1)) =
\nu\Gamma(y_1,y_2]$ where $u$ (resp. $l$) is the upper (resp.
lower) envelope, and the result follows.}

\pagebreak
\markboth{References}{References}
\printbibliography

\end{document}